%% file: RPP.tex
\documentclass[10pt]{iopart}

\usepackage{iopams}  % For AMS fonts
\usepackage{graphicx} % figures
\eqnobysec %equation numbering by section

\def\alm{a_{\ell m}}
\def\n{{\vec n}}
\newcommand\order[1] {${{\cal O}\! \left( #1 \right)}$}
\def\VEV#1{{\left\langle #1 \right\rangle}}
\graphicspath{{./}{Figures/}}
\begin{document}

\review[Data Analysis Methods for the Cosmic Microwave
Background]{Data Analysis Methods for the Cosmic Microwave Background}

\author{M. Tristram$^1$ and K. Ganga$^2$}

\address{
$^1$ LAL;
     Universit\'e Paris-Sud 11,
     B\^atiment 200,
     91400 Orsay,
     France
\\
$^2$ APC;
     10, rue Alice Domon et L\'eonie Duquet,
     75013 Paris,
     France
}
\ead{\mailto{tristram@lal.in2p3.fr}}

\begin{abstract}
  In this review, we give an overview of some of the major aspects of
  data reduction and analysis for the Cosmic Microwave Background.
  Since its prediction and discovery in the last century, the Cosmic
  Microwave Background Radiation has proven itself to be one of our
  most valuable tools for precision cosmology. Recently, and
  especially when combined with complementary cosmological data,
  measurements of the CMB anisotropies have provided us with a wealth
  of quantitive information about the birth, evolution, and structure
  of our Universe. We begin with a simple, general introduction to the
  physics of the CMB, including a basic overview of the experiments
  which take CMB data. The focus, however, will be the data analysis
  treatment of CMB data sets.
\end{abstract}

%Uncomment for PACS numbers title message
%\pacs{00.00, 20.00, 42.10}
% Keywords required only for MST, PB, PMB, PM, JOA, JOB? 
%\vspace{2pc}
%\noindent{\it Keywords}: Article preparation, IOP journals
% Uncomment for Submitted to journal title message
%\submitto{\JPA}
% Comment out if separate title page not required

\maketitle

%\tableofcontents
%\newpage

%%%%%%%%%%%%%%%%%%%%%%%%%%%%%%%%%%%%%%%%%%%%%%%%%%%%%%%%%%%%%%%%%
%  Introduction
%%%%%%%%%%%%%%%%%%%%%%%%%%%%%%%%%%%%%%%%%%%%%%%%%%%%%%%%%%%%%%%%%
\section{Introduction}

\subsection{History}

In 1964, Penzias and Wilson discovered a roughly 3.5~K noise excess
from the sky, using a communications antenna at Holmdel, New Jersey.
While serendipitous, this turned out to be a detection of the Cosmic
Microwave Background radiation (CMB), for which they were awarded the
Nobel Prize in 1978 \cite{PenziasWilson}.

In 1948, Alpher and Herman had published the idea that photons coming
from the primordial Universe could form a thermal bath at
approximatively 5~K \cite{alpher48}, while the present general
physical description of the CMB was obtained in the
'60s \cite{Dicke65}.

\subsection{CMB radiation}

In 1929, Edwin Hubble inferred that distant galaxies are moving away
from us with velocities roughly proportional to their
distance \cite{hubble29}. This is now considered the first evidence
for the expansion of our Universe. Given this expansion, we can assume
that the Universe was much denser and hotter earlier in its history.
Far enough back in time, the photons in the Universe would have had
enough energy to ionize hydrogen. Thus, we believe that sometime in
the past, the Universe would have consisted of a ``soup'' of
electrons, protons and photons, all in thermal equilibrium, coupled
electromagnetically via the equation~:
$$
e + p \leftrightarrows H + \gamma.
$$

Moving forward in time from this point, the Universe expands, and the
temperature decreases. The temperature will decrease to the point
where there are no longer appreciably many photons which can ionize
Hydrogen, so the protons and electrons will combine to form Hydrogen,
and stay in this form. This is called the epoch of recombination.  At
this point, the photons are no longer effectively coupled to the
charged particles, and they essentially travel unimpeded to this day.
This is the CMB we see today.

\subsection{A black body}

At the time of decoupling, constituents of the Universe are in thermal
equilibrium, so the electromagnetic spectrum of the CMB photons is a
black body, for which the intensity is
$$
I_\nu = {2h\nu^3\over c^2}{1\over e^{h\nu/k_BT} - 1}.
$$

\begin{figure}[htbp]
   \centering
    \includegraphics[angle=270,width=\textwidth]{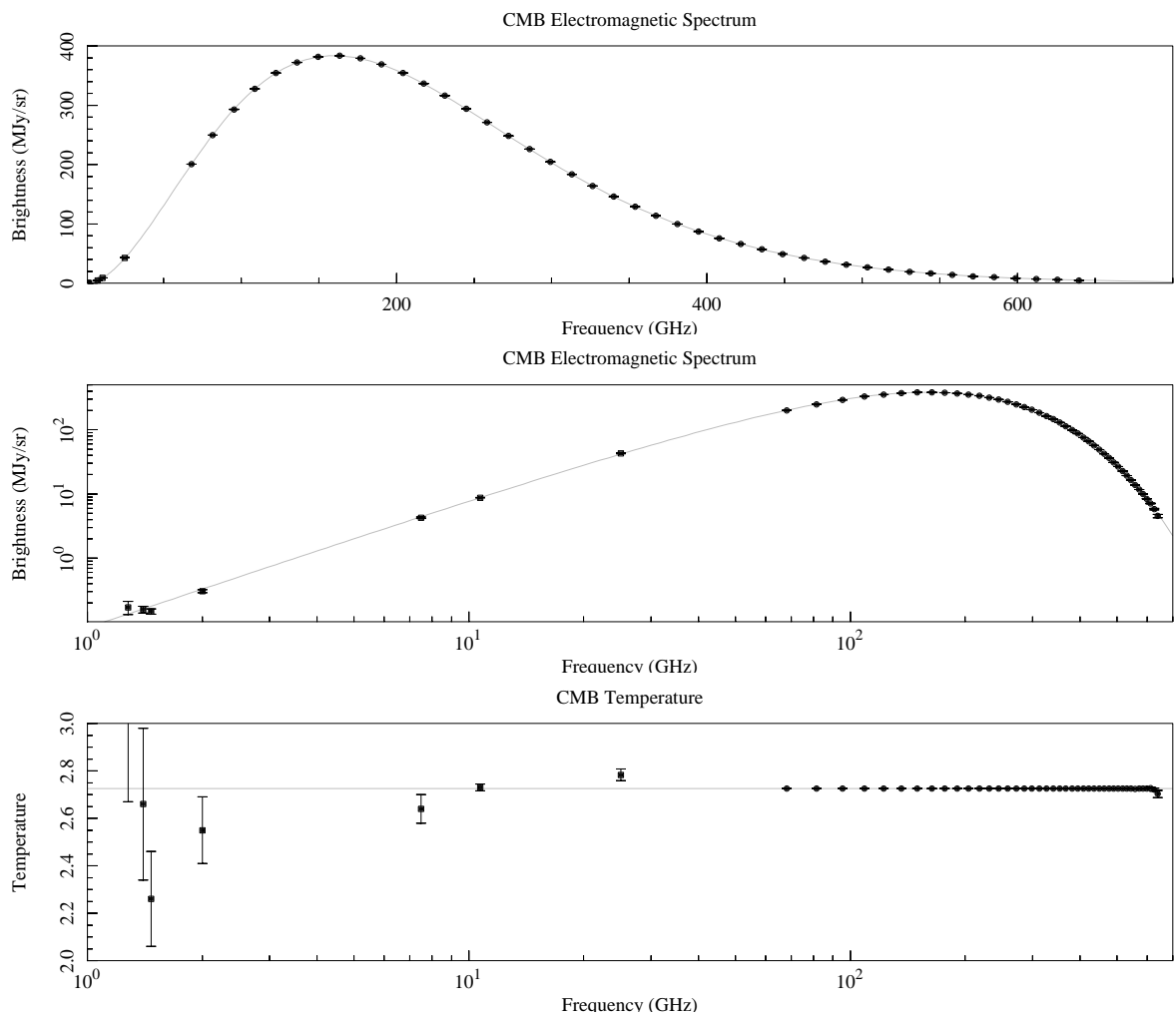}
    \caption{Electromagnetic spectrum of the CMB (top two panels) and
      measurements of the temperature of the CMB (bottom panel). The
      grey line indicates a blackbody with temperature 2.725~K. Error
      bars are included on all points, but in many cases are too small
      to discern. }
    \label{fig:firas}
\end{figure}

This prediction was verified by NASA in 1989 with the FIRAS instrument
on board the Cosmic Background Explorer (COBE) satellite. After a year
of observation, FIRAS measured a spectrum that was in near-perfect
agreement with the predictions (figure~\ref{fig:firas}). A recent
re-analysis of FIRAS data gives a black body temperature of
$2.725\pm0.001$~K (\cite{fixsen02}).

%%%%%%%%%%%%%%%%%%%%%%%
% FIG
% figure spectre de corps noir
%%%%%%%%%%%%%%%%%%%%%%% 

CMB practitioners often use the somewhat opaque ``CMB'', or
``thermodynamic'', units, assuming that small brightness variations
are related to small deviations in temperature from that of the CMB as
a whole.  Thus, these units can be obtained from the derivative of a
blackbody with respect to temperature via the equation:
$$
\Delta T
=
T_\mathrm{cmb}
\left(
  {2h\nu^3\over c^2}
  {e^{h\nu/kT_\mathrm{cmb}}\over \left(e^{h\nu/kT_\mathrm{cmb}}-1\right)^2}
  {h\nu\over kT_\mathrm{cmb}}
\right)^{-1}
\Delta B.
$$

\subsection{Dipole}

%%%%%%%%%%%%%%%%%%%%%%%
% FIG
% The dipole from COBE/DMR
\begin{figure}[htbp]
   \centering
    \includegraphics[width=\textwidth]{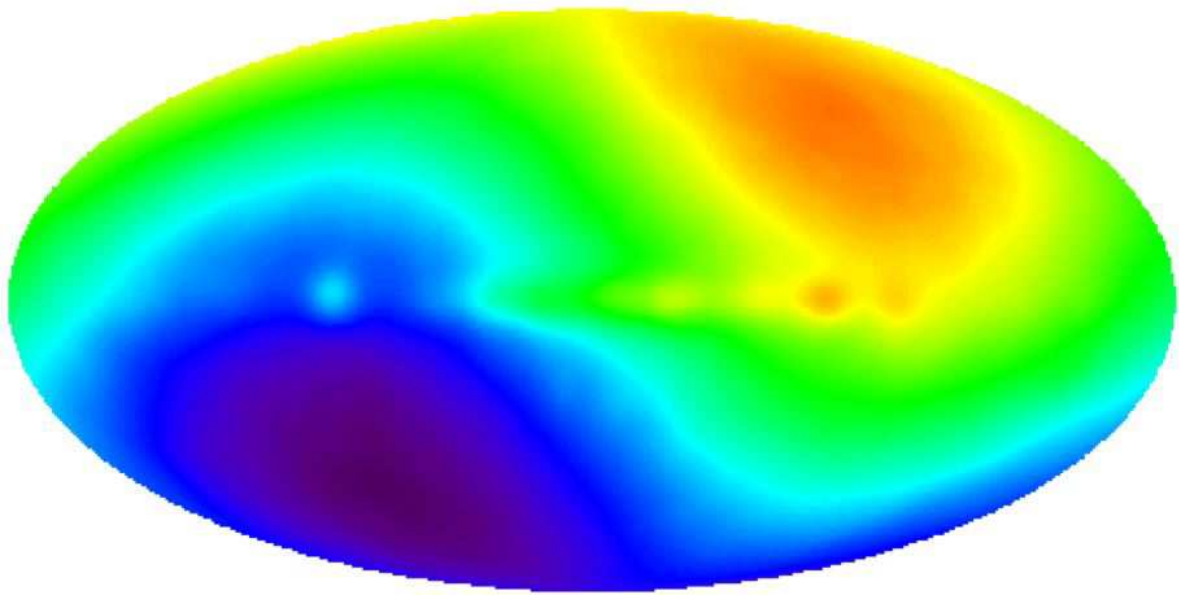}
    \caption{The CMB dipole as seen by COBE/DMR, from NASA's Legacy
      Archive for Microwave Background Data Analysis~(http://lambda.gsfc.nasa.gov).
      The overall blue-to-red variation indicates the CMB dipole. The
      faint features in the center of the map represent the plane of
      our Galaxy.}
    \label{fig:dipole}
\end{figure}
%%%%%%%%%%%%%%%%%%%%%%% 

After removing the mean value of the CMB, one finds a dipole pattern
with an amplitude of roughly 0.1\% of the average CMB temperature.
This is due to the doppler shift of CMB photons from the relative
motion of the Solar System with respect to the rest frame of the CMB.
CMB photons are seen as colder or hotter depending on the direction of
observation following, to first order
$$
\Delta T
=
T_\mathrm{cmb}\cdot\left({v\over c}\right)\cdot\cos\theta.
$$

COBE and WMAP \cite{hinshaw06} have measured the orientation and the
amplitude of the dipole (figure~\ref{fig:dipole}). To first order, it is well-described by
$$
\Delta T(\theta) = 3.358 \times 10^{-3} \cos \theta \mbox{~K},
$$
where $\theta$ is the angle between the direction of observation and
the dipole axis. The measured dipole implies that our Solar System is
traveling at roughly 370~km/s with respect to the rest frame of the
CMB. The motion of the Earth around the Sun contributes a roughly 10\%
modulation to this effect, which has been removed from this figure.

The motion of the Earth around the Sun also produces an additional
dipole contribution. This effect is another order of magnitude lower
than that of the dipole due to the motion of the Solar System with
respect to the CMB rest frame.  However, given that the dynamics of
the Earth within the Solar System are very well understood, this
signal provides a very convenient method to calibrate any CMB
anisotropy measurements, if an experiment can measure these large
scale variations.

\subsection{Primordial anisotropies}

As the Universe is expanding today, it must have been much smaller
earlier in its history. It must therefore have been much hotter,
meaning that both matter and the photons in the Universe had more
energy.  Imagine the point, very early in the history of the
Universe, where the photons each have much more energy than that
needed to ionize a Hydrogen atom. At this point, the matter and the
photons are in good thermal equilibrium. 

From this point, as the Universe cools, there are fewer and fewer
photons with enough energy to ionize hydrogen. At a certain point, the
mean free path of the photons becomes comparable to the size of the
accessible Universe, and the protons and electrons are essentially
free to combine permanently into hydrogen atoms. This period is given
the rather confusing name of ``recombination'' -- confusing since the
protons and electrons have never been consistently combined until this
point. From this point on, the photons are only lightly coupled to the
now neutral matter -- the Hydrogen atoms.

\begin{figure}[htbp]
  \centering
  \includegraphics[width=0.7\textwidth]{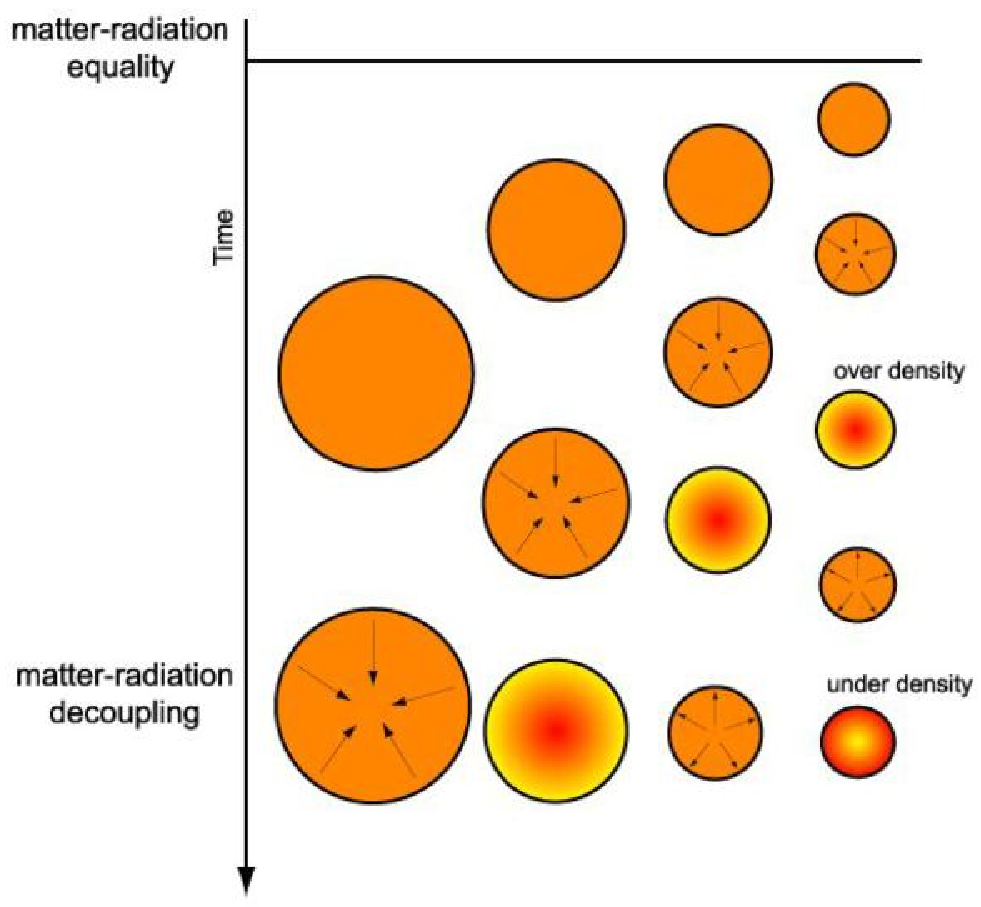}
  \includegraphics[width=0.7\textwidth]{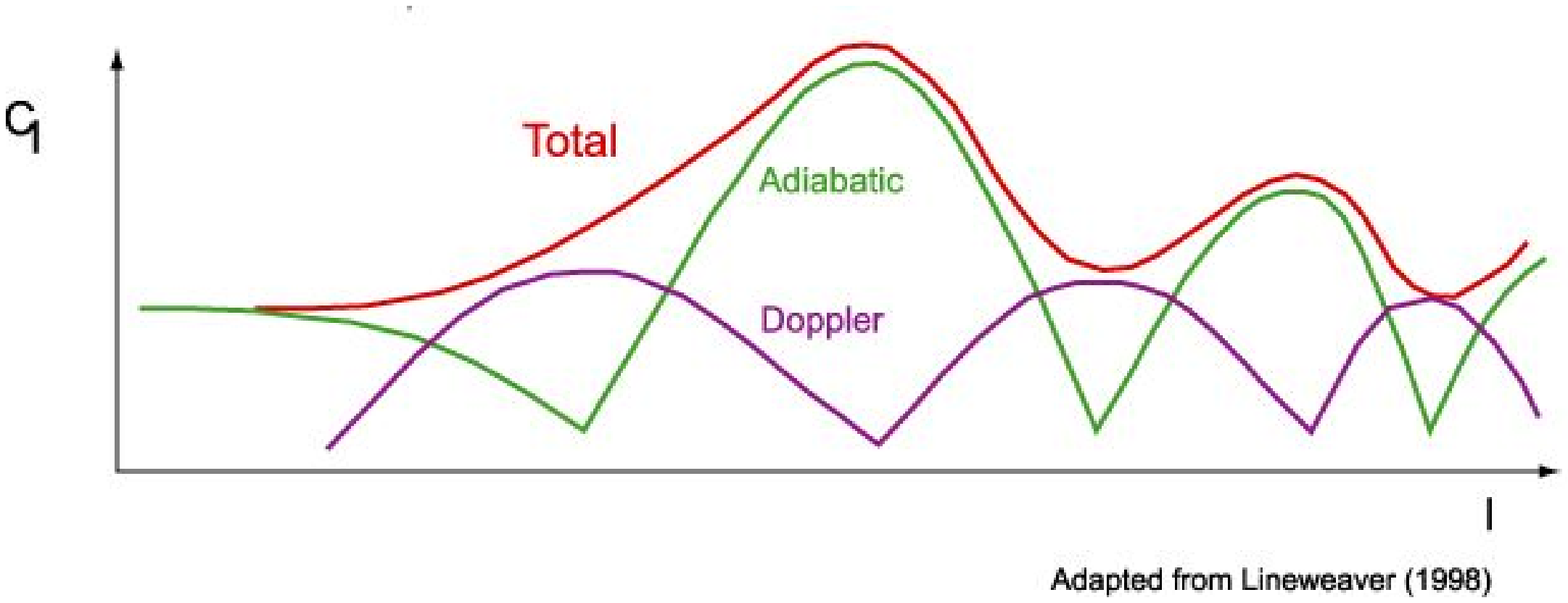}
  \caption{{\bf Acoustic oscillations and doppler peaks.}  Small
    structures come into the horizon earlier than larger ones and
    start oscillating. At the time of decoupling, we can observe the
    phase shifting of oscillations through the variation of amplitude
    fluctuation in temperature with respect to the size of the
    structures (characterized by the multipole $\ell$). 
    Figure adapted from \cite{lineweaver97}.}
  \label{fig:oscillations}
\end{figure}

While the process of cooling is happening, imagine a volume of this
photon-matter.  Gravitational instatibility, seeded by some small
deviation from uniformity, can cause the matter in the volume to
compress. However, photon pressure will tend to push such
overdensities apart. Thus, we are in a situation where structures of a
given size go through a series of compactifications and rarefactions.
Smaller regions will go through a series of compactifications and
rarefactions before recombination. Larger regions will do so less
often. These are called ``acoustic oscillations'' in the fluid. See
figure~\ref{fig:oscillations}

The so-called ``first peak'' in the power spectrum represents the
scale at which matter has just had time to maximally compress before
the recombination, which freezes these anisotropies into the photon
signature. This next peak represents the scale at which a single
compactification and a single rarefication has happened, etc.

Depending on which angular scales we are interested in, the primordial
anisotropies have amplitudes of roughly one part in 100000 of the CMB
mean. While quantitative estimation of the anisotropies caused by a
number of effects has been done, we give below a brief description of
a few of them.

%%%%%%%%%%%%%%%%%%%%%%%
% FIG
% figure de lineweaver, oscillation dans le spectre de puissance
%%%%%%%%%%%%%%%%%%%%%%% 

\subsubsection{Temperature:}

Depending on angular scales, one can describe three major effects
which cause anisotropies in the CMB:
\begin{itemize}
\item Adiabatic perturbations: Quantum fluctuations in the vacuum
  produce fluctuations of the density $\rho$. In inflation theories,
  these perturbations are adiabatic and Gaussian.  For a given density
  perturbation, the temperature fluctuation is
  $$
  \frac{\Delta T}{T} = \frac{1}{3} \frac{\delta \rho}{\rho}.
  $$
\item Gravitational perturbation (Sachs-Wolfe \cite{sachs-wolfe67}):
  When a photon falls into (or climbs out of) a gravitational well,
  its energy grows (or decreases) and it is thereby blueshifted (or
  redshifted).  Thus, on the sky, matter over-densities correspond to
  cold spots and under-densities correspond to hot spots. It must also
  be remembered that the Universe is expanding during this process so
  that when a photon traverses a gravitational potential change, the
  photon will see a different potential on entry and on exit of the
  well or hill.
\item Kinetic perturbation (Doppler): Variation of the primordial
  plasma velocities implies a Doppler effect on CMB photons. This
  shifting is proportional to the fluid velocity $v$, relative to the
  observer
  $$
  \frac{\Delta T}{T} \propto v.
  $$
  This effect vanishes along the line of sight for scales smaller than
  the depth of the last scattering surface, but can be seen on
  large scales.  
\end{itemize}

\subsubsection{Polarization:}

\begin{figure}[h!]
   \centering
    \includegraphics[width=0.65\textwidth]{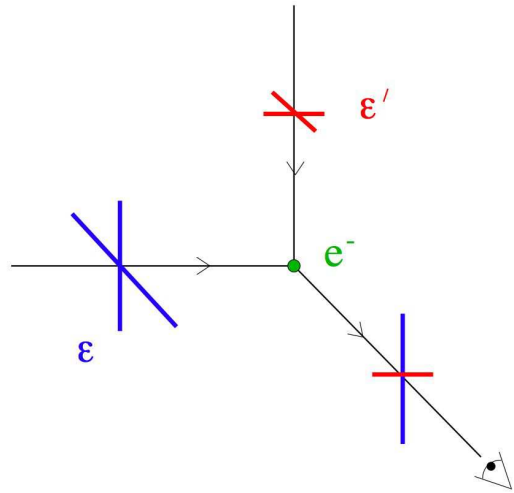}
    \caption{{\bf CMB polarization and quadrupolar anisotropies}.  A
      flux of photons with a quadrupolar anisotropy
      ($\epsilon$,$\epsilon'$) scatters from an electron resulting in
      linearly polarized radiation.}
    \label{fig:thomson}
\end{figure}

Polarization in the primordial CMB anisotropies comes from Thomson
scattering by electrons (figure~\ref{fig:thomson}). One can show via symmetry that only local
quadrupolar anisotropies of the radiation can produce a linear
polarization of the CMB photons. This is illustrated in the expression
of the differential Thomson scattering cross-section of a electron on
a non-polarized radiation
$$
\frac{d\sigma}{d\Omega} = \frac{3\sigma_T}{8\pi} |\epsilon \cdot
\epsilon'|^2.
$$

%%%%%%%%%%%%%%%%%%%%%%%
%FIG
%figure Thomson
%%%%%%%%%%%%%%%%%%%%%%% 

\begin{figure}[h!]
   \centering
    \includegraphics[scale=0.6]{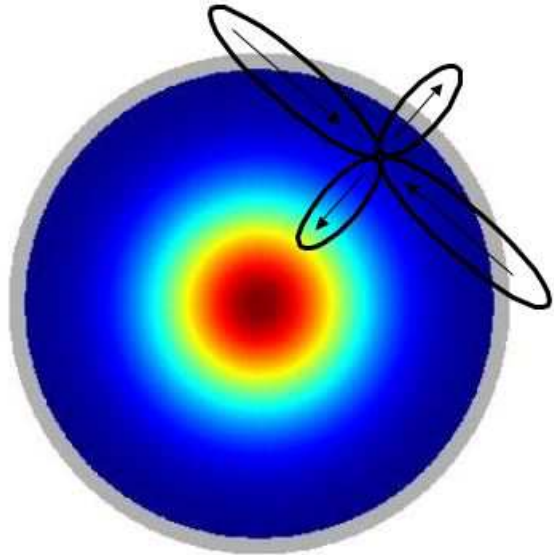}
    \caption{{\bf quadrupolar anisotropies formation} on a
      over-dense region. Electron along the over-density radius
      move away from each other whereas those that belong to a
      same density contour get closer.}
    \label{fig:quadrupole}
\end{figure}

Local quadrupolar anisotropies can be due to three different effects:
\begin{description}
\item[Scalar perturbations] scalar modes from density perturbations
  can cause quadrupolar anisotropies. See figure~\ref{fig:quadrupole}.
\item[Vector perturbations] Vortex movements of the primordial fluid
  can produce quadripolar anisotropies. They are not necessarily
  linked to an over-density. In most of inflationary models, these
  perturbations are negligible.
\item[Tensor perturbations] A gravitational wave passing through a
  density fluctuation can modify the shape of a gravitational well.  A
  symmetrical well become elliptical producing quadrupolar
  anisotropies.  
\end{description}

%%%%%%%%%%%%%%%%%%%%%%%
%FIG
%figure quadrupole
%%%%%%%%%%%%%%%%%%%%%%% 

\subsection{Secondary anisotropies}

Between the last scattering surface and our detectors, CMB photons can
encounter a number of perturbations. These produce so-called
``secondary'' anisotropies. They are usually either gravitational or
due to Compton scattering with electrons. The effects on the angular
power spectra are more fully described in \cite{hu95a}.

\subsubsection{Gravitational effects:}

\begin{description}
\item[Integrated Sachs-Wolfe:] This results from the variation of the
  gravitational field along the path of a photon as the Universe
  expands. This effect is limited. It can reach $\delta T/T \simeq
  10^{-6}$ at large angular scales.
\item[Gravitational lensing:] This is a distortion of the
  gravitational field due to massive objects (galaxies, clusters) that
  modify a photon's trajectory \cite{seljak00}. The angular power
  spectrum is smoothed by a few percent, which can make the small
  oscillations in the power spectrum at high multipoles disappear.
\item[Rees-Sciama:] \cite{rees-sciama}. This is linked to the
  development of gravitational wells with time. Photons that fall into
  a well need more energy to escape it than they received when
  entering; that is, the photons loose energy, if the well develops.
  This effect arises mostly when structures are forming. The {\em rms}
  amplitude of this effect is around $\delta T/T = 10^{-7}$ for a
  degree scale \cite{hu95a}. It can reach $\delta T/T \simeq 10^{-6}$
  for smaller scales (around 10 -- 40~arcmin) and can even become
  dominant below 40~arcsec \cite{seljak96a}.
\end{description}

\subsubsection{Scattering effects:}

\begin{description}
\item[Sunyaev-Zel'dovich (SZ) effect:] This is an inverse Compton
  effect, in which photons increase their energy by scattering from
  free electrons within hot gazes inside clusters \cite{zeldovich69},
  so that it is mostly significant at small angular scales. To first
  order, this slightly increases the energy of each photon and thus
  shifts the CMB electromagnetic spectrum. To second order, if the
  cluster is moving, one should also see a kinetic effect due to bulk
  motion of the cluster.  At large angular scales, the SZ effect can
  be seen due to diffuse scattering inside our own cluster.
  Anisotropies can reach $\delta T/T \simeq 10^{-4}$ for scales that
  range between a degree and arc-minutes.
\item[Reionization:] This corresponds to a period where the Universe
  becomes globally ionized once again, after
  recombination \cite{gunn65}.  During this period, free electrons
  will once again scattered CMB photons. It probably appears during
  structure formation ($z=6-20$). The effect on the CMB is visible
  both at small angular scales (suppressing the power from clusters)
  and at large scales.
\end{description}

\subsection{Foregrounds}
\label{sec:foregrounds}

CMB measurements can be contaminated by other astrophysical emissions
arising from our neighborhood \cite{bouchet99a}. Some examples are:

\begin{itemize}
\item{\bf Synchrotron emission.}  Relativistic electrons accelerated
  by a magnetic field produce synchrotron radiation, with a spectrum
  depending on both the intensity of the magnetic field and energy and
  flux of the electrons. The Galactic magnetic field of order a few nG
  is strong enough to produce this effect. The energy spectrum of the
  electrons is usually modeled as a power law, $\nu^{-\beta}$, with
  $\beta \simeq 3$ \cite{dezotti99}.  Synchrotron is the dominant
  foreground for for lower CMB frequency observations.
\item{\bf Bremsstrahlung (or \textit{free-free}) radiation.}  In a hot
  gas, ions decelerate free electrons, thereby producing thermal
  radiation. Once again, the {\it free-free} spectrum can often be
  modeled as a power law with spectral index $\beta \simeq 2.1$
  \cite{dezotti99}. As with synchrotron, {\it free-free} emission is
  most evident at lower CMB frequencies.
\item{\bf Galactic dust emission.}  Cold dust within our own Galaxy
  can emit via thermal radiation ({\it vibrational dust}) or by
  excitation of their electrical dipolar moment ({\it rotational
    dust}).  Thermal radiation is modelled as a grey body at $T \sim
  17$~K, with an emission maximum in the far-infrared. In the
  radio-millimetric domain, the dust emissivity can be modeled as
  $\nu^2$ \cite{schlegel98}. Vibrational dust emission has been
  claimed to have been seen between 10 and 100~GHz and with a maximum
  around 20~GHz \cite{watson05}, though there is still debate.
\item{\bf Extragalactic point sources.}  Some point sources can emit
  in the radio-millimetric domain. To avoid contamination by these,
  they are masked before the CMB power spectrum is estimated. For the
  background of undetected sources, their effect on the CMB spectrum
  is evaluated with Monte Carlos.
\end{itemize}

%%%%%%%%%%%%%%%%%%%%%%%
%FIG
%figure de spectre des avants-plans
\begin{figure}[htbp]
   \centering
    \includegraphics[width=12cm,height=6cm]{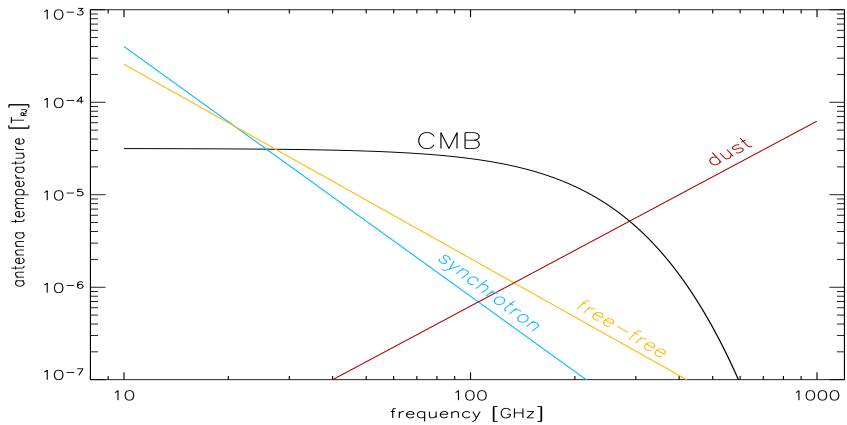}
    \caption{{\bf Foregrounds spectra} compared to CMB one ({\it
        black}). Amplitudes are normalized to the Sachs-Wolfe plateau.
      Synchrotron ({\it blue}) and {\it free-free} ({\it yellow}) are
      dominating at low frequency until $\sim$30~GHz. Dust ({\it red})
      is dominating at higher frequency (above 300~GHz).}
    \label{fig:foregrounds}
\end{figure}
%%%%%%%%%%%%%%%%%%%%%%% 

Figure~\ref{fig:foregrounds} shows representative foreground spectra,
though they may vary depending on location on the sky. CMB experiments
usually measure the CMB in the window between $\sim$20 and 300~GHz,
while measurements at higher and lower frequencies help estimate and
limit the level of foreground contaminants within a the CMB band.

While some experiments have measured polarized foregrounds, notably
Parkes \cite{giardino02} and WMAP \cite{barnes03} for synchrotron and
free-free and {\sc Archeops} for diffuse dust emission on large
angular scales \cite{benoit04,ponthieu05}, foreground polarization
over the full sky are still not well known.  Thus foreground residuals
have become one of the largest (if not the largest) source of
systematic errors in CMB analyses.

\subsection{Angular power spectra}

To describe CMB anisotropies, we decompose both temperature and
polarization sky maps into spherical harmonics coefficients. Most
inflationary models predict fluctuations that give gaussian
anisotropies in the linear regime \cite{hu97,linde99,liddle00}. In
such cases, the angular power spectra both in temperature and
polarization contain all the cosmological information of CMB.

\subsubsection{Temperature:}

The spherical harmonics, $Y_{\ell m}$, form an orthogonal basis
defined on the sphere. The decomposition of a scalar map into
spherical harmonic coefficients $a_{\ell m}^T$ reads
$$
\frac{\Delta T\left(\n\right)}{T} 
= 
\sum_{\ell=0}^{\infty} \sum_{m=-\ell}^{\ell} 
\alm^T Y_{\ell m}\left(\n\right),
$$
where $\alm^T$ satisfy
$$
a_{\ell m}^T 
= 
\int \frac{\Delta T\left(\n\right)}{T} Y_{\ell m}^*(\n) d\n.
$$

The multipole $\ell$ represent the inverse of the angular scale.
We can define the angular power spectrum $C_\ell^T$ by
$$
C_\ell^T  = \left< | \alm^T |^2 \right>
$$

Moreover, for gaussian anisotropies, the $\alm$ distribution is also
gaussian and its variance is the angular power spectrum $C_\ell$:
\begin{eqnarray}
  \VEV{a_{\ell m}} 
  & = & 
  0,
  \nonumber 
  \\
  \VEV{a_{\ell m} a_{\ell' m'}} 
  & = & 
  C_\ell \delta_{\ell\ell'} \delta_{m m'}.
  \nonumber
\end{eqnarray}

Thus we can write an estimator $\tilde{C}_\ell^T$ of the power
spectrum that reads
$$
\tilde{C}_\ell^T 
= 
\frac{1}{2\ell +1} \sum_{m=-\ell}^\ell \alm^T \alm^{T*}.
$$

\begin{figure}[htbp]
   \centering
    \includegraphics[width=\textwidth]{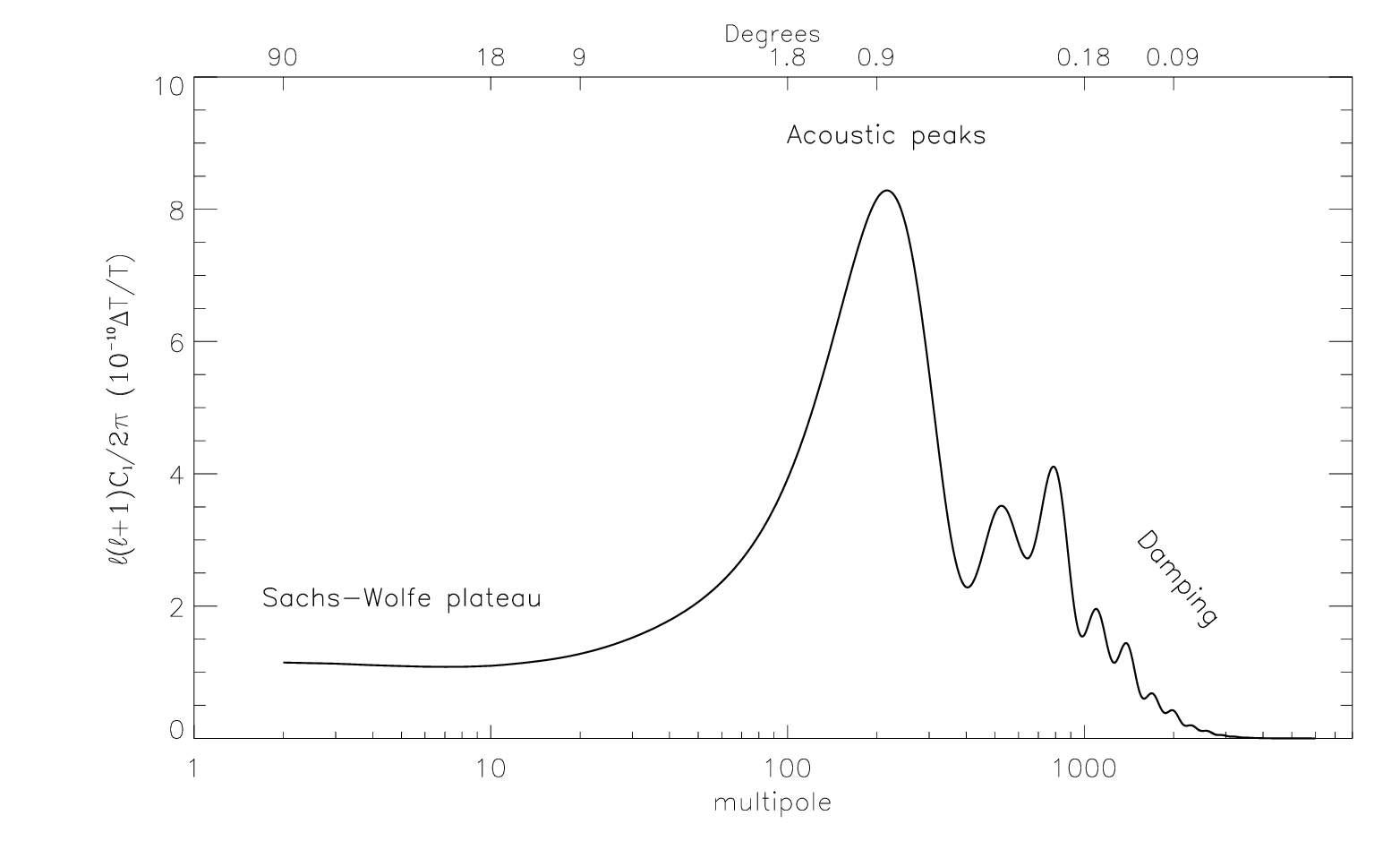}
    \caption{{\bf Temperature angular power spectrum}: power on the
      sky as a function of the multipole $\ell$ or angular scales. We
      can distinguish three regions from left to right: the
      Sachs-Wolfe plateau, the acoustic peaks and the damping region.}
    \label{fig:cell}
\end{figure}

The angular power spectrum in temperature shows three distinct regions
(see figure~\ref{fig:cell}):

\begin{enumerate}
\item {\bf The Sachs-Wolfe plateau.}  For scales larger than the
  horizon, causality dictates that fluctuations never evolve.
  Anisotropies come from initial fluctuations of photons (form the
  gravitational field) and from the Sachs-Wolfe
  effect \cite{sachs-wolfe67}. Since the spectrum of the fluctuations
  from the gravitational field is scale invariant, the temperature
  fluctuations are statistically identical and the angular power
  spectrum is nearly flat at large scales (small multipole $\ell$).
\item {\bf Acoustic oscillations.}  For scales smaller than the
  horizon, in a matter dominated Universe, the fluid undergoes
  acoustic oscillations that are adiabatic. Baryons fall into
  gravitational wells, whereas photon radiation pushes them apart.
  This induces acoustic oscillations of matter which imprints on the
  photons. Structures enter the horizon progressively (starting with
  the smallest ones) resulting in a progression of oscillations
  depending on the size of the structures
  (fig~\ref{fig:oscillations}). Peaks in the angular power spectrum
  reflect these phase-differences for scales smaller than the horizon
  ($\ell \gtrsim 180$). Differences in the electron velocities at the
  time of the last scattering also imply a second order Doppler
  effect.
\item {\bf Damping region.}  At still smaller angular scales, the
  spectrum is damped, mainly due to residual diffusion of photons,
  which smooths structures with scales smaller than the mean free path
  (Silk damping). Furthermore, the recombination process is not
  instantaneous, with a finite width resulting in a more gradual
  damping. 
\end{enumerate}

%%%%%%%%%%%%%%%%%%%%%%%
%FIG
%figure quadrupole
%%%%%%%%%%%%%%%%%%%%%%% 

The correlation function of the signal on the sky is
$$
C\left(\theta\right)
=
\sum_{\ell}{2\ell+1\over 4\pi}C_\ell P_\ell\left(\cos\theta\right)
$$

We note that the power spectra are often plotted as 
$$
f\left(C_l\right)
=
{l\left(l+1\right)\over 2\pi}\cdot C_l. 
$$
This is convenient, as when shown like this the area under the curve
is roughly equal to the variance of the signal on the sky.

\subsubsection{Polarisation:}

The Stokes formalism allows one to describe polarized radiation with
four scalars: I, Q, U and V. For a polarized wave propagating along
the $z$ axis, Stokes parameters are
$$
\begin{array}{rcl}
  I & = & \VEV{ | E_x |^2 + |E_y|^2},         \\
  Q & = & \VEV{ | E_x |^2 - |E_y|^2},         \\
  U & = & \VEV{2 Re(E_xE_y^*)},\ \mathrm{and} \\
  V & = & \VEV{2 Im(E_xE_y^*)}.
\end{array}
$$

Unpolarized light is described by $Q = U = V = 0$.  $Q$ and $U$
characterize the linear polarization for the photon whereas $V$
describes the circular polarization. $I$ and $V$ are rotation
invariant whereas $Q$ and $U$ depend on the frame of reference..
Conservation of the total energy of a wave implies that
$$
I^2 \ge Q^2 + U^2 + V^2.
$$

Stokes parameters can be summed for a superposition of incoherent
waves.  Thomson diffusion cannot create circular polarization as it
does not modify the phases but only the amplitudes of each component.
Thus, for CMB, $V = 0$.

In the same way as for temperature, we can define polarized angular
power spectra using the decomposition in spherical harmonics for the
$Q$ and $U$ parameters on the sky.  To do this, we use the scalar $E$
and pseudo-scalar $B$ quantities defined from Stokes parameters but
which are independent from the frame of reference. The decomposition
is made using the spin-two harmonics:
$$
(Q \pm iU)(\n) 
= 
\sum_{\ell m} a_{\pm2\ell m} \ _{\pm2} Y_{\ell}^{m}(\n).
$$
The connection between $Q/U$ and $E/B$ in spherical harmonic space is
\begin{eqnarray}
  a_{\ell m}^E & = & - \frac{a_{2\ell m} + a_{-2\ell m}}{2} \nonumber \\
  a_{\ell m}^B & = & i \frac{a_{2\ell m} - a_{-2\ell m}}{2} \nonumber,
\end{eqnarray}
where we can define the purely polarized angular power spectra
$C_\ell^E$ and $C_\ell^B$ as
\begin{eqnarray}
  C_\ell^E & = & \VEV{ | \alm^E |^2 } \nonumber \\
  C_\ell^B & = & \VEV{ | \alm^B |^2 } \nonumber.
\end{eqnarray}

Polarization of the CMB is due to quadrupolar anisotropies at the last
scattering surface. We thus expect correlations between temperature
anisotropies and polarized anisotropies, which can be described by the
temperature-polarization angular cross-power spectra
\begin{eqnarray}
  C_\ell^{TE} & = & \VEV{ \alm^T \alm^{E*} } \nonumber \\
  C_\ell^{TB} & = & \VEV{ \alm^T \alm^{B*} } \nonumber.
\end{eqnarray}

Finally, second-order spin spherical harmonics properties implies that
$$
C_\ell^{EB} = \VEV{ \alm^E \alm^{B*} } = 0.
$$

One can demonstrate that for scalar perturbations $E \ne 0$ and $B=0$,
whereas for tensor perturbations $E,B \ne 0$.  Thus detecting $B$
polarization in the CMB could indicate the presence of tensor modes
and thus be an indication of gravitational waves.

%%%%%%%%%%%%%%%%%%%%%%% 
% FIG figure spectres de puissance
\begin{figure}[htpb]
   \centering
    \includegraphics[width=\textwidth]{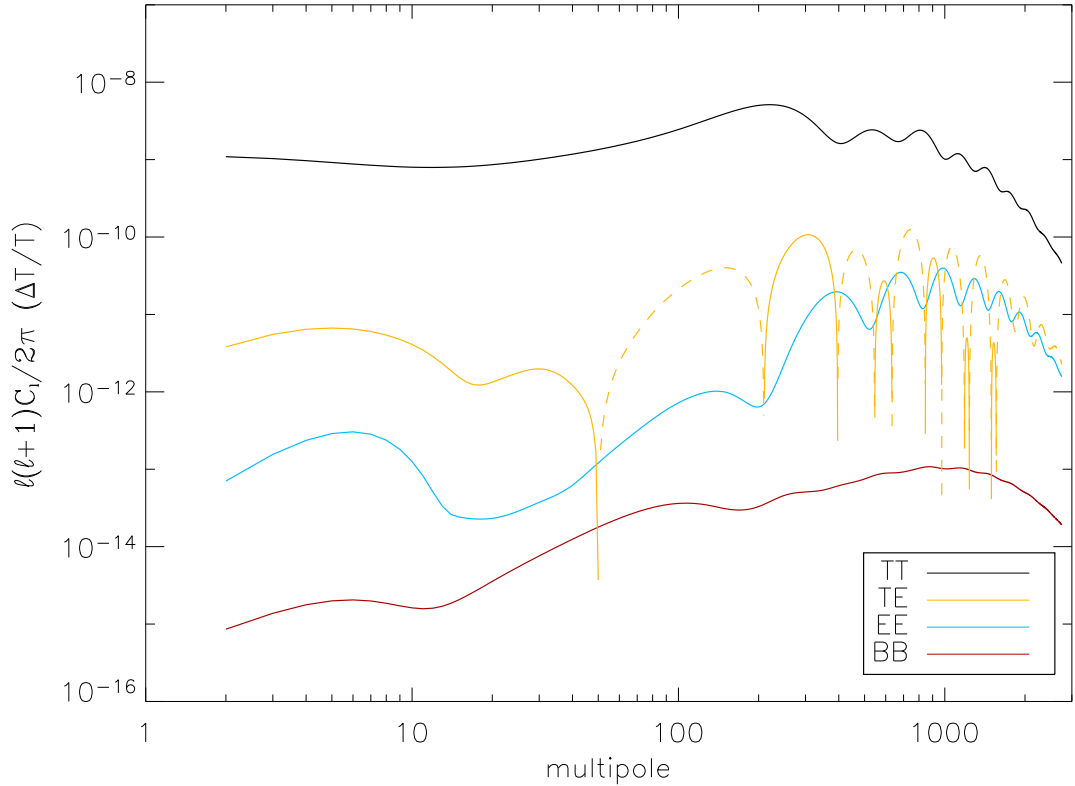}
    \caption{{\bf Angular power spectra} both in temperature,
      polarization and cross temperature-polarization. {\it From top
        to bottom~:} temperature $TT$, cross-spectrum $TE$, purely
      polarized spectrum $E$ et an optimistic estimation of the $B$
      spectrum.}
    \label{fig:all_cl}
\end{figure}
%%%%%%%%%%%%%%%%%%%%%%% 

As with the temperature spectra, polarized spectra also show peaks for
scales smaller than approximately a degree (see
figure~\ref{fig:all_cl}).  These are sharper for polarization, as they
are due to velocity gradients of the photon-baryon fluid at the time
of decoupling only (Doppler oscillation). Consequently, they are
shifted by $\pi/2$ with respect to temperature peaks, which are
dominated by density fluctuations. The correlation between the two is
characterized by the cross temperature-polarization power spectrum
$C_\ell^{TE}$.  The $E$ spectrum is at the level of a few percent of
the temperature spectrum. The amplitude of the $B$ modes are still
unknown but should be at least one or two order of magnitude below
that of the $E$ modes. On large angular scales, the $B$-mode amplitude
is strongly linked to the energy of inflation
$E_{inf}$ \cite{zaldarriaga02}:
$$
\left[ \ell(\ell +1)/2\pi \right] C_\ell^B 
\simeq 
0.024^2 \left( E_{inf}/10^{16} \right)^4 \mu\mbox{K}^2.
$$

For gaussian fluctuations, we can also define the estimators for each
spectrum:
\begin{eqnarray}
  \tilde{C}_\ell^E 
  & = & 
  \frac{1}{2\ell + 1} \sum_{m=-\ell}^{\ell} | a_{\ell m}^E |^2 
  \nonumber \\
  \tilde{C}_\ell^B 
  & = & 
  \frac{1}{2\ell + 1} \sum_{m=-\ell}^{\ell} | a_{\ell m}^B |^2 
  \nonumber \\
  \tilde{C}_\ell^{TE} 
  & = & 
  \frac{1}{2\ell + 1} \sum_{m=-\ell}^{\ell} a_{\ell m}^T a_{\ell m}^{E*} 
  \nonumber \\
  \tilde{C}_\ell^{TB} 
  & = & 
  \frac{1}{2\ell + 1} \sum_{m=-\ell}^{\ell} a_{\ell m}^T a_{\ell m}^{B*}.
  \nonumber
\end{eqnarray}

\subsubsection{Cosmic and sample variance:}

For both temperature and polarization, the $\alm$ coefficients are
gaussian distributed with a mean of zero and a variance given by the
$C_\ell$.  Each of these coefficients has $2\ell+1$ degrees of
freedom, corresponding to the $2\ell+1$ $m$-values for a given $\ell$, 
due to the fact that we can only measure a single realization of our Universe 
from one location. This induce an intrinsic variance on the estimated $C_\ell$, 
called cosmic variance that is equal to
$$
Var_{cosmic}(\tilde{C}_\ell) = \frac{2}{2\ell+1} C^2_\ell .
$$
Note that for large angular scales, this can become significant. 

Moreover, CMB anisotropies measurements cannot cover the whole sky.
Even for satellites, foreground emission residuals can be comparable
to the CMB signal and we therefore must use a mask that reduces the
effective coverage. For each multipole, the number of degrees of
freedom increase as a function inverse of the observed area $f_{sky}$
and so the associated variance (called sample variance)
$$
Var_{sample}(\tilde{C}_\ell) = \frac{2}{(2\ell+1)f_{sky}} C^2_\ell.
$$

%%%%%%%%%%%%%%%%%%%%%%%%%%%%%%%%%%%%%%%%%%%%%%%%%%%%%%%%%%%%%%%%%
%  Observation technics
%%%%%%%%%%%%%%%%%%%%%%%%%%%%%%%%%%%%%%%%%%%%%%%%%%%%%%%%%%%%%%%%%
\section{Instruments}

\subsection{Observation sites}

CMB experiments have observed the CMB from a variety of different
sites; from telescopes sited all over the globe, to balloons, to
satellites in Earth orbits, and now even to satellites at the second
Sun-Earth Lagrange point.  Each site has its own advantages and
disadvantages. Specifically:

\begin{itemize}
\item Sky coverage: Full-sky coverage is usually only achieved by
  satellites, which have the unique combination of long observation
  times and unobstraucted views of the sky. Balloon-borne experiments
  can cover a significant fractions of the sky (such as $\sim$30\% for
  Archeops \cite{benoit03} and FIRS \cite{ganga93}). The balloon-borne
  19~GHz experiment \cite{boughn92} covered almost the full sky by
  making multiple flights from North American and Australia, but
  balloon observations are often limited to much smaller regions (for
  example, $<$10\% of the sky for BOOMERanG or 0.25\% for MAXIMA
  \cite{rabii06}). Ground base measurements can usually only cover a
  few percent of the sky.
\item Resolution: Detector resolution is directly linked to the size
  of the telescope and the wavelength of observation. Satellites and
  balloons are thus usually limited in resolution compared to
  ground-based measurements due to weight constraints.
\item Atmosphere: Satellite, obviously, do not have problems with
  terrestrial atmosphere. Ground-based measurements, on the other
  hand, are hampered by atmospheric emissions such as water vapor,
  which absorbs microwave radiation.  Thus ground base telescopes for
  the CMB are operated from dry, high altitude locations such as the
  Chilean Andes or the South Pole.  Balloon experiments, flying at
  tens of kilometers from the ground, offer a compromise.
  Nevertheless, there are still some atmospheric effects from, for
  example, ozone clouds.
\item Observing time: Balloon-borne CMB experiments have usually been
  single-night observations, though some experiments have had multiple
  flights, and so-called ``long duration'' flights of over a week are
  now becoming common. Satellite experiments have observed for a
  number of years. Ground-based experiments have also observed for
  years.
\end{itemize}

\subsection{Scanning strategy}

With a given amount of observing time, which is often limited by site
conditions or resources, a CMB experiment's scanning strategy aims to:
\begin{itemize}
  \item minimize foreground contributions
  \item provide the redundancy necessary to analyze noise and other
    unforeseen effects.
  \item provide the best possible calibration and instrumental
    characterization. E.g., quasars for pointing reconstruction. 
  \item minimize atmospheric effects
  \item allow a decent measurement of the power spectrum. 
\end{itemize}

\subsubsection{Foregrounds:}

As noted in section~\ref{sec:foregrounds}, foreground emission can be
a major contaminant to CMB measurements. Thus, all CMB experiments
take care to either avoid observing regions with excessive foreground
emission, or to reject these regions when the data are analyzed.

To this point, all satellite-based CMB experiments have used scanning
strategies which covered the entire sky, motivated by a combination of
technical simplicity, and the fact that it is one of the few ways to
consistently measure the largest scale anisotropies in the CMB.
However, this means that some regions, such as the Galactic plane, are
not useable, and must be excised from the data. Balloon experiments
such as the 19~GHz Experiment \cite{boughn92}, FIRS \cite{ganga93} and
Archeops \cite{benoit03} have been used to make large fractions of
the sky. In these experiments, the Galactic plane is treated in much
the same way as for satellites, with data in high-foreground regions
simply avoided in the analyses.

\begin{figure}[htbp]
   \centering
    \includegraphics[width=\textwidth]{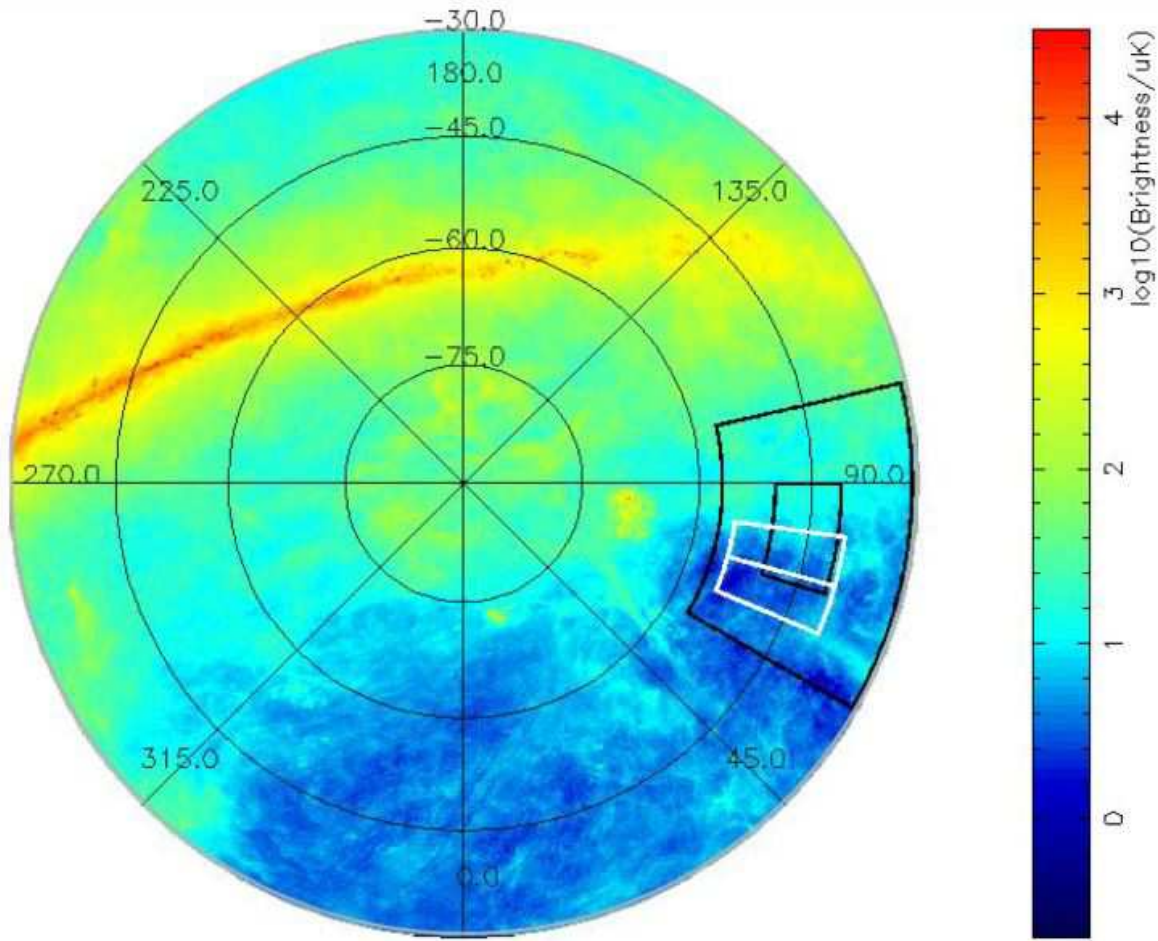}
    \caption{ This grapic shows an estimate of the emission from dust
      which would be seen by polarization-sensitive experiments at
      150~GHz. The center of the plot is the zenith at the South Pole
      -- that is, declination -90~degrees. The edge of the plot is
      -30~degrees (that is, it shows the ``bottom'' of the celestial 
      sphere). Zero right ascension is down in the plot,
      increasing in the counter-clockwise direction. The brightness
      estimates come from application of a model by Finkbeiner et
      al. \cite{finkbeiner99}. The boxes to the right represent the
      sky coverage for QUaD and BOOMERanG, two CMB anisotropy
      experiments. The larger and smaller black boxes are the
      BOOMERanG so-called ``shallow'' and ``deep'' fields,
      respectively. The two white boxes are the two fields QUaD
      observed in their first season of
      observations \cite{hinderks05}.}
    \label{fig:QUaDBoomCover}
\end{figure}

A number of balloon-borne experiments, however, have been used to make
maps of localized regions of a few percent of the sky. In such cases,
the observation fields are chosen to coincide with low emission from
our Galaxy. In addition, almost all ground-based experiments map but a
few percent of the sky at most, and use the same foreground avoidance
technique. An example of this is shown in figure~\ref{fig:QUaDBoomCover}.

%%%%%%%%%%%%%%%%%%%%%%%
% FIG
%%%%%%%%%%%%%%%%%%%%%%%

\subsubsection{Redundancy:}

It is notoriously difficult to keep sensitive experiments stable for
long periods of time, and these experiments are no exception.
Temperature and atmospheric changes, as well as a host of other
experimental possibilities conspire to allow the baselines, or zeros
of these experiments to change. If only a single measurement were made
of each point on the sky, it would be quite difficult to differentiate
between sky signals and so-called ``systematic'' effects. It would
also be difficult to differentiate between ``real'' signal and
``random noise''.

To this end, experiments endeavor to observe a given part of the sky
in as many different ways as possible. It is highly desirable to
observe all points measured on as many different time scales as
possible, from as many different directions as possible.

Note that while it is desireable to observe a given spot in as many
different directions as possible, there are often overriding concerns.
As an example, we note that most ground-based and many balloon
experiments try to observe without changing the elevation of
observations, since changing elevation with change the column depth of
atmosphere through which the experiment is observing and will thus
change loading and equilibrium of the experiment. Often an experiment
can get ``cross-linking'' in various scans, since as the Earth rotates
a given scanning will ``rotate'' on the sky. There are, however, a
number of experiments which are making, or have made, observations
from the South Pole. From this unique vantage point (along with the
North Pole, though there have not been any CMB experiments fielded
there for obvious reasons), as the earth rotates, one cannot change
declinations except by changing elevations. Thus, these experiments
live without the benefits of cross-linking. From the South pole,
however, it is often noted that the atmosphere is low enough that
experiments can work without it.

\subsubsection{Calibration:}

While it is possible to calibrate an instrument using special
techniques, by far the most accepted procedure is to use astronomical
sources to calibrate, preferable sources which can be seen as part of
the routine observations done of the CMB. In this way the experiment
is calibrated in the configuration used to make the cosmological
measurements themselves, and assumptions or extrapolations between the
``routine'' measurements and the calibration measurements need not be
made. 

The most desirable source to use would be something with the frequency
spectrum of the CMB anisotropies themselves. While an increasing
number of experiments are using the CMB anisotropies themselves, as
measured by previous experiments, to calibrate, a number of
experiments have also used the CMB dipole, which also has the same
spectrum. When doing this, care must be taken to account for the
roughly 10\% variation in the dipole due to the motion of the Earth
around the Sun.

For experiments which do not cover a large enough area to use the
dipole for calibration, the scanning strategy will ideally cover a
planet or some bright, well-known point source which, along with
understanding of the beam and bandpass of the instrument, can provide
a flux calibration. In addition, these sources can be used to refine
pointing and beam models.

\subsubsection{Power Spectrum Sampling:}

Different regions of the power represent structures of different sizes
-- lower multipoles representing structure at larger angular scales
and higher multipoles representing structure at smaller angular
scales. For experiments interested in measuring the structure on the
largest scales, the scanning strategy must, of course, cover areas of
these sizes. In addition, in order to have sufficient statistics, the
experiment will usually have to cover a number of patches of the size
of interest, in order to integrate down the ``sample
variance'' \cite{scott94}, the inherent variance we will find from one
patch of a given size to another, even when the fluctuations in both
are given by the same underlying model.

In addition, if one fails to observe large enough regions, even if one
can formally measure power spectrum values for a given multipole,
without enough observations the spectrum points at different
multipoles will be {\em correlated}, effectively limiting the
experiments resolution in multipole space \cite{tegmark96b}.

\subsection{Detectors}

\subsubsection{Low Frequency:}

Here, ``Low Frequency'' refers to frequencies between roughly 15 and
95~GHz.

From the COBE Differential Microwave Radiometer (DMR) and the
Wilkinson Microwave Anisotropy Probe (WMAP), to the low frequency
instrument (LFI) of Planck, we can see three examples of low frequency
radiometers.

A radiometer is a device whose output voltage is proportional to the
power received by a horn antenna.  The output is then sent to an
amplifier such as a High Electron Mobility Transistor (HEMT).  For
radiometers sensitive to polarization, one can use an OrthoMode
Transducer (OMT) to separate the orthogonal polarizations with minimal
losses and cross-talk. The two orthogonal linear polarizations are
then directed into separate amplifiers.

The radiometer equation \cite{dicke46}
$$
\delta T 
= 
T_{sys} \sqrt{ \frac{1}{\Delta\nu \tau} + \left(\frac{\Delta G}{G} \right)^2 },
$$
gives the total power radiometer's sensitivity for an integrating
period $\tau$, a frequency-dependent power response $G(\nu)$, an input
referenced system noise temperature $T_{sys}$ and the effective RF
bandwidth $\Delta \nu = \left[ \int G(\nu) d\nu \right]^2 / \int
G^2(\nu)d\nu$. The second term represents the noise coming from the
gain variation of the radiometer during the integration time $\tau$.

Due to their low noise and wide bandwidth, HEMTs are good candidates
for measurements of the CMB.  Unfortunately, these amplifiers exhibit
long scale variations of their gain that limit sensitivity of the
radiometers.  Reducing 1/f noise can be done using differential
radiometers. That is, by switching the inputs from two antennas or an
antenna and a reference load, the temperature of which is close to the
measured signal (as for Planck-LFI). In the first case, difference
signal is then constructed using two orthogonally polarized channels.
In the second case, a hybrid coupler can provide two phase-switch
signals from the reference load and the sky signal. In both cases,
switching enhances the instrument's performances in two ways: (1)
since both signals are amplified by the same chains, gain fluctuations
in either amplifier chain act identically on both signals so that
common mode gain fluctuations cancel; (2) the phase switches introduce
a 180$^\circ$ relative phase change between two signal paths. Thus,
low frequency (1/f) noise is common mode and vanishes.

These low frequency radiometers are usually cooled to lower than
100~K, which reduces amplifier noise and makes them more sensitive.

The DMR was launched in 1989. It detected structure in the CMB angular
distribution at angular scales $\gtrsim 7^\circ$ \cite{smoot90}, using
two Dicke-switched radiometers at frequencies: 31, 53 and
90~GHz, with noise temperature of 250, 80 and 60 times the quantum
limit respectively, fed by pairs of feed horns pointed at the sky.

WMAP \cite{jarosik03} was launched in June of 2001 and is currently
observing the sky in five frequency bands: 23, 33, 41, 61 and
94~GHz, with arrays of radiatively-cooled radiometers fed by a
differential two-telescope optical system. Radiometer noise
temperatures are 15--25 times the quantum limit, with angular
resolution ranging from 56~arcmin to 14~arcmin.

The LFI instrument, slated to fly on the Planck
satellite \cite{bersanelli00}, with its array of cryogenically cooled
radiometers, represents another advance in the state of the art. It is
designed to produce images of the sky (including polarized components)
at 30, 44 and 70~GHz, with high sensitivity.

%%%%%%%%%%%%%%%%%%%%%%%
% FIG
\begin{figure}[htbp]
   \centering
    \includegraphics[width=\textwidth]{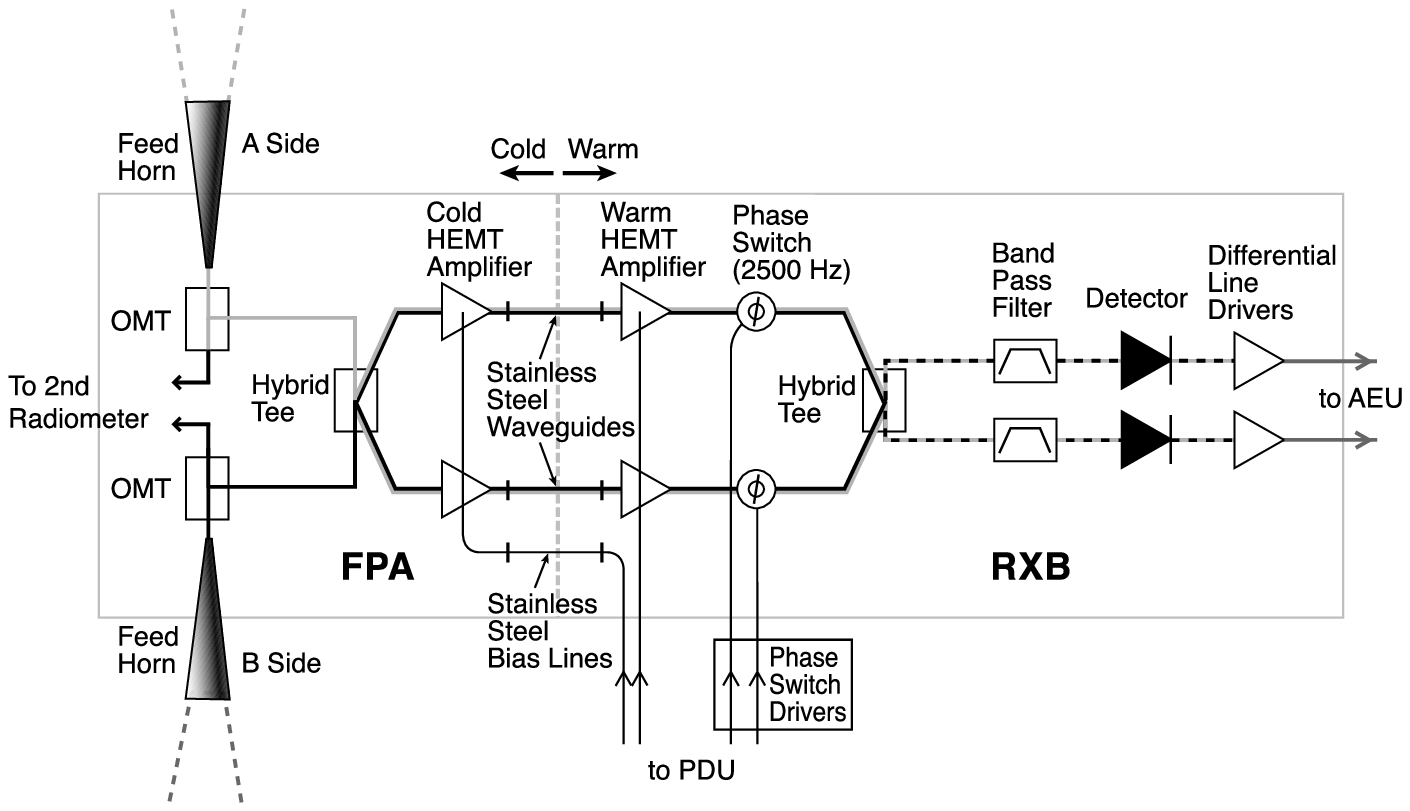}
    \caption{ Layout of an individual WMAP radiometer. Components on
      the cold ({\it left}) side of the stainless steel waveguides are
      located in the FPA and are passively cooled to 90 K in flight.
      (figure extracted from \cite{jarosik03})}
    \label{fig:radiometer}
\end{figure}
%%%%%%%%%%%%%%%%%%%%%%% 

\subsubsection{High Frequency:}

Here, ``High Frequency'' refers to frequencies between roughly 95 and
250~GHz. 

Bolometric detectors \cite{chanin83} are micro-fabricated devices in
which the incoming radiation is absorbed by a grid, causing an
increase in temperature. This temperature increase is measured by a
Neutron Transmutation Doped (NTD) germanium thermistor, which provides
high sensitivity with sufficient stability. These detectors give
extremely high performance, yet are insensitive to ionizing radiation
and microphonic effects.

\begin{figure}[htbp]
   \centering
    \includegraphics[height=6.5cm]{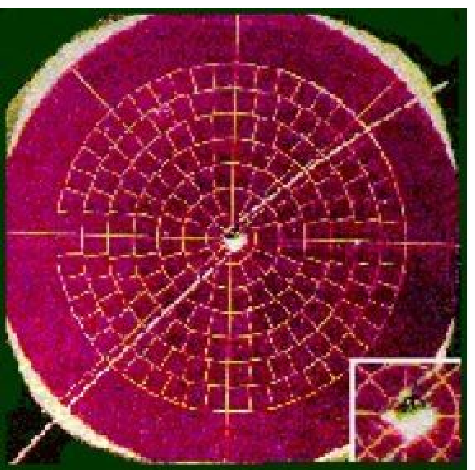}
    \hspace{0.5cm}
    \includegraphics[height=6.5cm]{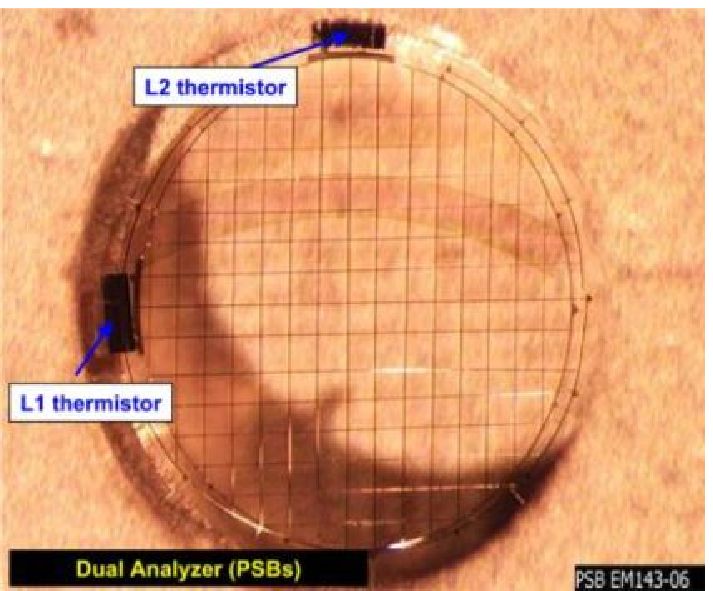}
    \caption{ Spider-Web Bolometer ({\it left}) and
      Polarization-Sensitive Bolometer ({\it right}) from Planck
      High Frequency Instrument.}
    \label{fig:bolometer}
\end{figure}

The modern ``total power'' CMB bolometers are grids which resemble
spider webs (figure~\ref{fig:bolometer}), with characteristic scales related to the wavelength of
the radiation of interst, reducing background coming from lower
wavelengths.  Moreover, this configuration enhances sensitivity and
reduce the time response and cross-section with particles. Its lower
mass gets him less sensitive to vibration.  For Polarization-Sensitive
Bolometers (PSB), radiation is absorbed by two orthogonal grids of
parallel resistive wires, each of which absorbs only the polarized
component with electric field parallel to the wires \cite{jones03}.
Polarized sky can be reconstructed using several detector
measurements.

%%%%%%%%%%%%%%%%%%%%%%%
%FIG
%%%%%%%%%%%%%%%%%%%%%%% 
Radiation from the telescope is coupled to the bolometer via horns and
filters to select the wavelength.

%%%%%%%%%%%%%%%%%%%%%%%
%FIG
\begin{figure}[htbp]
   \centering
    \includegraphics[width=\textwidth]{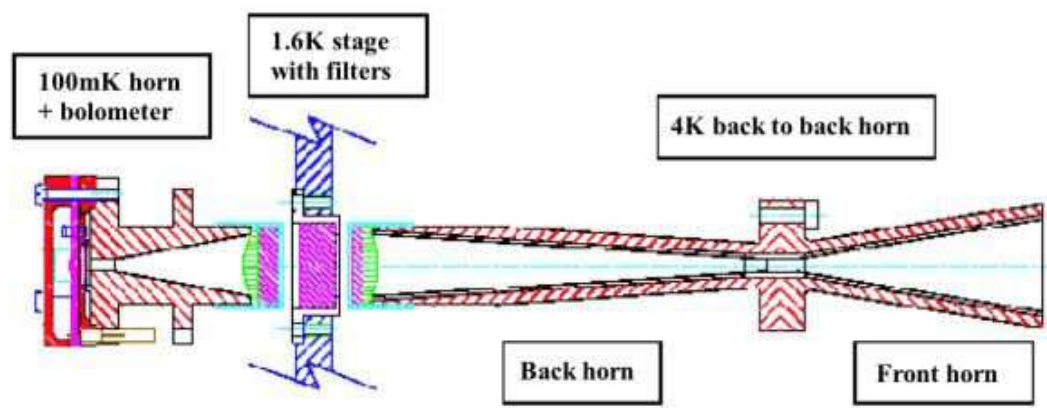}
    \caption{ Optical configuration for a single photometric pixel
      from Archeops or Planck focal plane.}
    \label{fig:photometric_pixel}
\end{figure}
%%%%%%%%%%%%%%%%%%%%%%% 

Thermodynamic sources of noise in a bolometer are coming from~:
\begin{enumerate}
\item phonon noise proportional to temperature;
\item Johnson noise linked to fluctuations of tension applied on the thermistor; 
\item photon noise coming from quantic nature of the incoming radiation.
\end{enumerate}
The fundamental limit of the sensitivity of a bolometer is phonon noise in the thermal
link between the absorber and the heat sink. In this case, the noise
equivalent power reads
$$
NEP = \gamma \sqrt{4 k_B T^2 G},
$$
where $G$ is the thermal conductance, $T$ the temperature of the bath,
and $\gamma$ takes into account the contribution from Johnson noise in
the NTD Ge thermistor.  For a given background load $Q$, maximum
sensitivity is achieved for $G \sim Q/T$ \cite{mather84}.  The time
constant of the bolometer is defined by $\tau = C/G$ where $C$ is the
heat capacity of the bolometer. The time constant is fixed by the
modulation scheme, putting a limit on the thermal conductance $G$.

Cooling down the system reduces the two first sources of noise such
that intrinsic photon noise become dominant.  Other sources of noise
are microphonic noise coming from vibrations and 1/f noise due to low
thermal drifts.

Using bolometers impose a cryogenic system to cool down detectors.
BOOMERanG \cite{crill03} uses a cryostat that operate at 270~mK.
Archeops \cite{benoit02} and Planck use a dilution cryostat that
insure 100~mK on the focal plane.

\subsection{Environment effects}

\subsubsection{Thermal effects:}

Detectors used for accurate measurements of temperature variations
such as CMB anisotropies are very sensitive to thermal variation of he
environment.  Thus experiments are designed to minimize the effects of
thermal variations across the focal plane and electronics which might
induce changes in the gain and offsets of the detectors. The
observatory environment is designed to be as stable as possible given
other the other constraints of the observations.  Satellites are now
placed at the second Sun-Earth Lagrange point, placing the Earth
between the Sun and the payload.  Moreover, the focal plane is looking
in the opposite direction from the Sun and large baffle prevent from
most of scattering light that could enter the instrument.  Balloon
also used scanning strategy that avoid Sun or fly during the arctic
night.  Ground-based experiments prevent with the Sun light using
baffles and operates during the night.

Anyway, thermal effects, that represent the largest source of
systematics at low frequency, are monitored using thermometers that
are used in the data analysis. The latter can also be used to regulate
some of the cryogenic stages.

\subsubsection{Electrical effects:}

Variation of electrical signal can affect the signal even for stable
thermal environment. These variations can be due to, for example,
solar flares, RF noise or voltage fluctuation.

Signal from a detector can also be related to another one via
electrical cross-talk that can be due to nonideal behavior of
electronics or pickup in the wiring hardness.

Usually, tests are made at ground before the observing period to
search for some parasitic effects.

\subsection{Interferometers}

The study of CMB anisotropies using interferometers goes back over two
decade. Nowadays, high resolution measurements of the CMB power
spectra have been made by ground based interferometers VSA
\cite{dickinson04}, DASI \cite{leitch05}, CBI \cite{rajguru05,
  readhead04}.

In contrast with thermometers that measure the total or differential
power, an interferometer directly measures the power spectrum of the
sky. Images of the sky can then be reconstructed using aperture
analysis.  They can cover continuously a large range of the power
spectrum since their angular resolution is determined by the number of
fields observed.  Moreover, the detection of only correlated signals
made them very stable to systematics such that ground pickup and
atmospheric emission.

An interferometer measures the average over a time long (compared to
the wavelength) of the electric fields vectors $E_1$ and $E_2$ of two
telescopes pointing on the same direction of the sky~:
$\VEV{E_1E_2^*}$.  For a monochromatic wave in the Fraunhofer limit,
the average $\VEV{E_1E_2^*}$ is the intensity times a phase factor.
The phase factor is given by the geometric path difference between the
source and the two telescopes in units of the wavelength.  When
integrating over the source plane, we obtain the visibility
$V(\mathbf{u})$ which is the Fourier Transform of the temperature
fluctuation on the sky $\Delta T(\hat{x})$ multiplied by the
instrument beam $B(\hat{x})$ \cite{tompson86}. The visibility reads
$$
V(\mathbf{u}) \propto \int d\hat{x} B(\hat{x}) \Delta T(\hat{x})
e^{2\pi i \mathbf{u} \cdot \hat{x}}
$$
where $\hat{x}$ is a unit pointing three-vector, $\mathbf{u}$ is the
conjugate variable characterizing the inverse angle measured in
wavelength.

The size of the aperture function $A(\hat{x})$ gives the size of
the map which means the coverage sky.  The maximum spacing determines
the resolution.  Considering the relatively small field of view of
interferometers, we can assume the small-angle approximation and treat
the sky as flat. In such conditions, for $u \gtrsim 10$ and $\ell
\gtrsim 60$, one can demonstrate that the visibility can be linked to
the angular power spectrum as
$$
u^2 S(u) 
\simeq 
\left. \frac{\ell(\ell+1)}{(2\pi)^2} C_\ell \right|_{\ell=2\pi u} 
$$

As we have seen, data analysis for interferometers is very specific
and we will not go into details in this review. For more complete
description, you can refer to \cite{martin88}, \cite{subrahmanyan93},
\cite{hobson96}, \cite{white99} or \cite{park03}.

%%%%%%%%%%%%%%%%%%%%%%%%%%%%%%%%%%%%%%%%%%%%%%%%%%%%%%%%%%%%%%%%%
%  Preproc
%%%%%%%%%%%%%%%%%%%%%%%%%%%%%%%%%%%%%%%%%%%%%%%%%%%%%%%%%%%%%%%%%
\section{Preprocessing}

\begin{figure}[htbp]
   \centering
    \includegraphics[height=9cm,width=\textwidth]{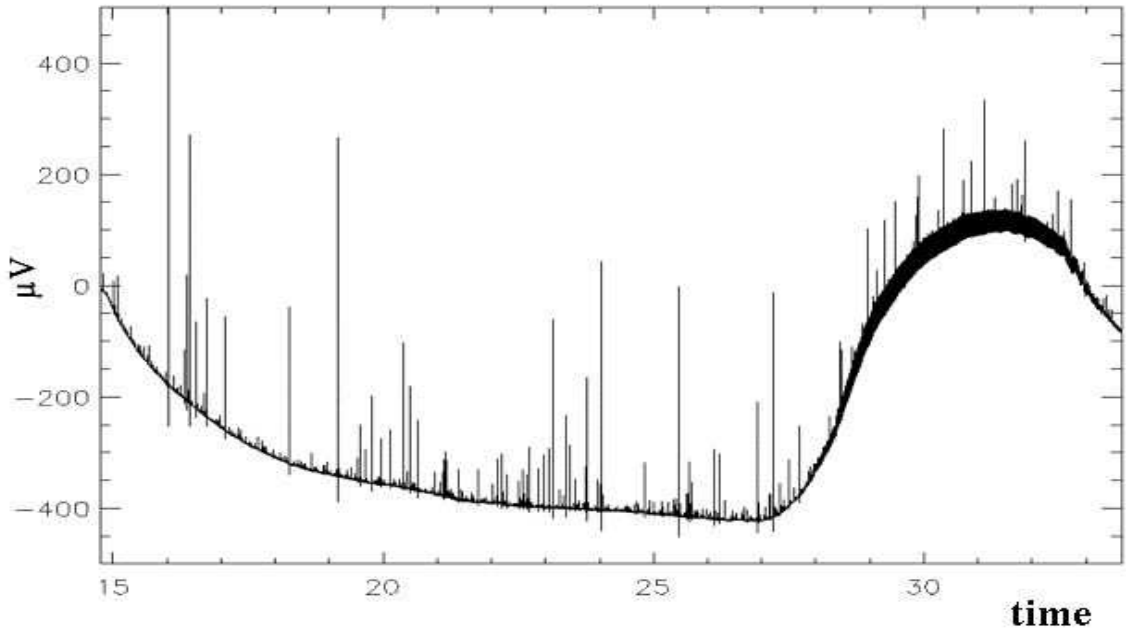}
    \caption{ Raw data of Archeops last flight for a bolometer at 143~GHz in arbitrary units.
      The slow drift is due to a slow change in temperature
      during the flight. The Sun rose after $\sim$12 hour of flight. Cosmic rays are
      clearly visible as spikes in the data. }
    \label{fig:rawdata}
\end{figure}

These steps are very instrument dependent.  From a general point of
view, we transform raw data (figure~\ref{fig:rawdata}) into a timeline or time-ordered data
(TOD).  More than just collecting data, this first step often deals
with decompression and demodulation data, as well as removing any
parasitic signals introduced by, for example, the readout electronics.
It may also correct for any non-linear response from the detectors and
may flag bad data.

\subsection{Demodulation}

Data from detectors (scientific signal) and thermometers (housekeeping
data) are often modulated in order to provide a method to lockin on
the signal. An AC square wave modulated bias, for example, transforms
the data into a series of alternative positive and negative values. In
the Fourier power spectrum of the data, this induces a peak in the
spectrum. This peak dominates the signal and needs to be removed for
demodulation.  This can be performed by filtering the data with a
low-pass filter considering the following constraints:
\begin{itemize}
\item the transition after the cut-off frequency must be sharp for
  complete removal of the modulation signal,
\item the ringing of the Fourier representation of the filter above
  the cut-off frequency needs to be below the approximately 2\% level,
  to avoid aliasing.
\end{itemize}
The cut-off frequency must be chosen below the Nyquist frequency and
above the cut-off due to both the beam pattern and the detector time
response in order to preserve the signal.

\subsection{Readout electronic noise}

When data is stored, it can be compressed into blocks before
recording.  The data recording can be delayed and a few data blocks
are buffered before recording. Small offset variations in the
electronics lead to significant differences between the mean value of
previously acquired blocks and those following, which induces a
parasitic signal on the data.  This parasitic signal shows up as
periodic pattern in frequency proportional to the ratio of the
acquisition frequency over the size of the block, depending on the
number of blocks buffered.

\subsection{Data flagging}

Raw data often contains periods that are suspected or known to be
unusable. It can be due to the absence of data or data dominated by
parasitic sources such as glitches, noise bursts or jumps due to
reconfigurations of the detectors. Those samples are flagged and could
be (for some specific purpose such as Fourier transform) filled by
constrained realizations of noise. Flagged data are simply not used to
make the final CMB maps and power spectra.

Methods to identify these effects are often based on iterative
detection of spikes before flagging. At each step, data can be
band-pass filtered or convolved with a specific template in order to
make the parasitic effect more visible.

Changes in detectors parameters such as the bias produce jumps in the
data. Microphonic noise coming from mechanical vibrations or sudden
releases of internal mechanical stress can also induce glitches.  In
addition, bolometers are also sensitive to cosmic ray hits. These are
therefore major sources of glitches in bolometer data.

The cosmic ray glitch rate depends on the effective surface of the
detector absorber and the observation site (ground, $\sim 1$ per hour,
or balloon/satellite, $\sim$ 1 per minute for a 1mm$^2$ detector
surface). The signature of cosmic-ray hits a delta function convolved
with the instrumental response. Thus, collections of cosmic ray
responses can be used to estimate the transfer function of the
detector and the electronics (see
section~\ref{sec:transfer_function}). Moreover, a model of energy
deposit taking into account the time response of the detectors and
electronics gives the shape of the signal as a function of time and
helps to estimate how long the data is badly affected by a cosmic ray
hit. An approximate model often used is
$$
g\left(t, t_i\right) 
= 
A\cdot e^{-\left(t-t_i\right)/\tau} \ast f_{em}~+~f_{base},
$$
where $\ast$ represents convolution, $A$ is the response amplitude,
$f_{base}$ is the baseline, $f_{em}$ is the electronic modulation
function and $\tau$ is the detector relaxation time constant. Some
detectors can show more complex transfer functions with several time
constants \cite{macias07,crill01} that can be related to where the
particles deposit their energy on detector.

This process might flag bright sources as cosmic rays. To avoid this,
detected glitches are compared with data taken at the same point on
the sky at another time with the same or some other detector to
confirm that the large signal isn't actually a strong signal on the
sky.

Housekeeping data from instrument can also be used to locate and flag specific bad
data such as repointing or changes of instrument parameters which can produce jumps.

The main objective is to flag parasitic signal above the noise level.
At the end of the process, a small fraction of the data is flagged
(usually less than a few percent).

%%%%%%%%%%%%%%%%%%%%%%%
%FIG
%%%%%%%%%%%%%%%%%%%%%%% 

%%%%%%%%%%%%%%%%%%%%%%%%%%%%%%%%%%%%%%%%%%%%%%%%%%%%%%%%%%%%%%%%%
%  Systematics
%%%%%%%%%%%%%%%%%%%%%%%%%%%%%%%%%%%%%%%%%%%%%%%%%%%%%%%%%%%%%%%%%

\section{Description and subtraction of systematics}

In this section we describe systematic effects that can be found in
CMB data analysis as well as the methods and algorithms used for their
subtraction.

\subsection{Description}

In a scanning instrument, multipoles of the CMB anisotropies are
encoded in time-ordered data at frequencies $f$ which depend on the
elevation $e$ and the sky scan speed $\theta'$ as
$$
f \simeq \frac{\theta'}{2\pi} \cos\left(e\ell\right).
$$

Depending on the scan strategy, we can define three distinct regimes
in the Fourier domain.
\begin{itemize}
\item First, the very low frequency components are mainly due to
  $1/f$-like noise both from detectors (for bolometers) and
  electronics (both for bolometers and radiometers). Long time-scale
  drifts from temperature changes of the cryogenic stages and the
  telescope can also be clearly seen in the time domain. For balloon
  experiments, drifts can also come from variation of air mass during
  the flight due to changes in the balloon altitude. Such systematics
  are highly correlated within detectors and can be monitored by
  housekeeping data from thermometers and altitude measurements.
\item Second, scan-synchroneous systematics are the most difficult to
  handle. Indeed, at the scan frequency and its harmonics, in addition
  to the CMB and other extraterrestrial emission, other components can
  be present experimental contamination can be present.
\item Finally the high frequency components are dominated by detector
  noise. At high frequencies, the Fourier spectrum is nearly flat.
  For bolometers, time response is closely described by a first order
  low-pass filter that cuts drastically high frequencies.  Microphonic
  noise can also put imprints on the high frequency noise.

\end{itemize}

\subsection{Subtraction}

Systematics that are monitored can be removed via a decorrelation
analysis using templates based on housekeeping data and external
and/or internal data.

Templates for atmospheric effects can be constructed using altitude
and elevation of the payload and a model of the atmosphere, or by
using higher frequency detectors which measure at a frequency where
atmospheric emission dominates over CMB and other emission. Blind
detectors (which are identical to the standard detectors but which
have been sealed off from light) are used as microphonic noise
monitors. Temperature measurements of different parts of the
experiment give us a handle on long-term drift temperature variations.

Correlation coefficients can be computed via linear regression before
the templates are subtracted to the data \cite{masi06, macias07}.
Templates and/or data can be filtered or smoothed depending on the
range of frequency of interest.

Although this decorrelation procedure is very efficient, one can often
still see correlated low frequency parasitic signals in detectors,
which creates stripes in the maps made. To avoid the mixing of the
detector signals at this stage of the processing of the data, this
effect is usually considered later, when the maps are made.

\subsection{Filtering and baseline removing}

The easiest way to get rid of long-term drifts, microphonics, or other
effects which are localized in Fourier space is simply to apply a
highpass, bandpass or a ``prewithening'' filter.

The purpose of the filter is to clean the data so that the
pixel-to-pixel covariance matrix (and thus the noise covariance of the
angular power spectrum) becomes simpler. But the filter should modify
the underlying signal as little as possible. Thus, noise properties
need to be checked after the data treatment and filtering could need
iteration.

Data from WMAP radiometers shows some $1/f$ noise at very low
frequency ($f_{knee}$ typically of a few mHz \cite{hinshaw03}). Even
though the effects are small relative to the white noise, it would
generate weak stripes of correlated noise along the scan paths. In
order to minimize these effect on the final maps, a prewhitening,
high-pass filtering procedure has been applied to the data. The method
is based on fitting a baseline to the TOD after removal of an
estimated sky signal. The baseline is subtracted before the signal is
added back in.

For experiments that perform large circles on the sky, the CMB signal
in the Fourier domain is located around the scan frequency, so that it
is negligible at higher frequencies. A low-pass filter can be used to
remove high frequency microphonic noise, while a high-pass filter can
be applied to remove very low frequency where $1/f$ noise dominates.

%%%%%%%%%%%%%%%%%%%%%%%%%%%%%%%%%%%%%%%%%%%%%%%%%%%%%%%%%%%%%%%%%
%  Pointing
%%%%%%%%%%%%%%%%%%%%%%%%%%%%%%%%%%%%%%%%%%%%%%%%%%%%%%%%%%%%%%%%%
\section{Pointing reconstruction}

Pointing reconstruction consists of determining for each sample where
the detectors are pointing in the sky.  The accurate {\it a
  posteriori} reconstruction is critical for mapping correctly the sky
signal.

\subsection{Method}

The first step is to reconstruct the pointing direction of the
telescope as a whole.  This is usually performed using a stellar or
solar sensor aligned with the direction of the telescope. This can be
combined with several attitude sensors measuring either absolute
angles (GPS) or angular velocities (gyroscopes).  This step can be
described mathematically as a rotational matrix, called the attitude
matrix, which converts from an Earth-based reference frame to the
telescope frame. It is defined by three Euler angles and so usually
described by a quaternion. From stellar sensor data we can reconstruct
the pointing direction by comparing observations to catalogs. The
sensor outputs can then combined using a Kalman
filter \cite{kalman60}, which recursively estimates the state of this
dynamic system from a series of incomplete and noisy measurements.

The positions of individual detectors with respect to the telescope
can then be reconstructed from measurements of bright, compact
sources, such as planets, or bright Galactic or extra-Galactic
sources.

\subsection{Accuracy}

The effect of an unknown, random pointing reconstruction error can be
modeled using the modified formula for the uncertainty in a power
spectrum mesaurement  \cite{knox95}:
$$
\frac{\Delta C_\ell}{C_\ell} 
= 
\sqrt{ \frac{2}{(2\ell+1)f_{sky}} } \left(1 + \frac{w}{C_\ell W_\ell}\right),
$$
where $w$ is the noise per beam, $f_{sky}$ is the fraction of the sky
covered, and $W_\ell$ is the transfer function of the beam ($W_\ell =
e^{-\ell(\ell+1)\sigma^2}$ for a gaussian beam). The beam causes a
loss of sensitivity at higher multipoles $\ell$. Pointing uncertainty
can be modeled as a smearing of the beam, which increasing the
effective beam width. Thus, pointing requirements for CMB experiments
are usually fixed by comparison with the level of noise at high
multipoles.

As an example, detail on methods for the pointing reconstruction of
Planck can be found in \cite{harrison04}.

\subsection{Focal plane reconstruction}

\begin{figure}[htbp]
   \centering
    \includegraphics[height=12cm,width=12cm]{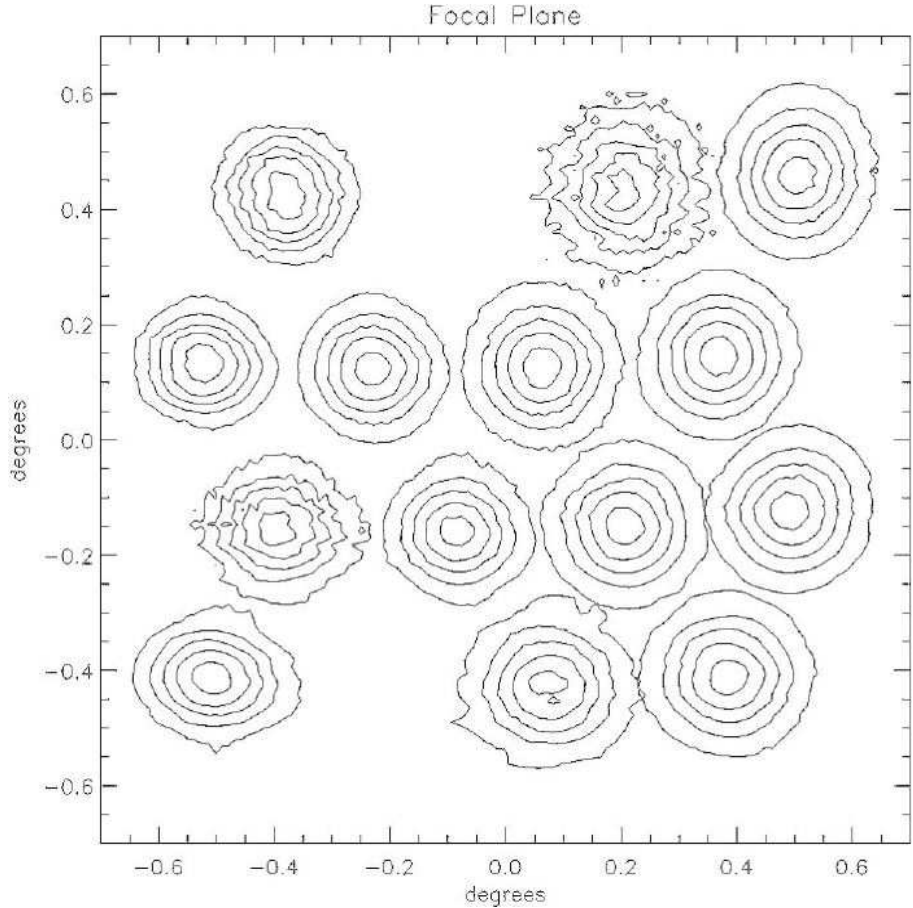}
    \caption{The MAXIMA-II focal plane. The contours, from the center
      of each beam out, represent the 90\%, 70\%, 50\%, 30\%, and 10\%
      levels respectively. (figure extracted from \cite{rabii06})}
    \label{fig:focal_plane}
\end{figure}

The position of each photometric pixel in the focal plane relative to
the Focal Plane Center is computed using a point source as reference (figure~\ref{fig:focal_plane}).
This then allows us to build the pointing of each detector using the
pointing reconstruction.

%%%%%%%%%%%%%%%%%%%%%%%
% FIG
%%%%%%%%%%%%%%%%%%%%%%% 

%%%%%%%%%%%%%%%%%%%%%%%%%%%%%%%%%%%%%%%%%%%%%%%%%%%%%%%%%%%%%%%%%
%  Detector response
%%%%%%%%%%%%%%%%%%%%%%%%%%%%%%%%%%%%%%%%%%%%%%%%%%%%%%%%%%%%%%%%%
\section{Detector response} 

This section describe the reconstruction of the focal plane
parameters: the time response of the detectors, the optical response
of the photometric pixels and the focal plane geometry on the
celestial sphere.

\subsection{Time response}
\label{sec:transfer_function}

The transfer function of the experiment is usually parameterized by a
thermal time constant of the detector and the properties of the
readout electronics and filters. The time response of the detectors
can often be described by a simple thermal model where the relaxation
follows $e^{-t/\tau}$.

The time constant $\tau$ can be evaluated on bright sources profiles.
Nevertheless, depending on the scanning strategy, it can be difficult
to disentangle from, or even degenerate with, the beam shape. This is
especially the case for experiments that scan the sky in one direction
only, with quasi-constant rotation speed (such as Archeops or Planck),
whereas scanning small patches back and forth allows one to deduce the
true shape of the beam below the leak due to the time constant (such
as for BOOMERanG or MAXIMA).

\begin{figure}[htbp]
   \centering
    \includegraphics[width=\textwidth]{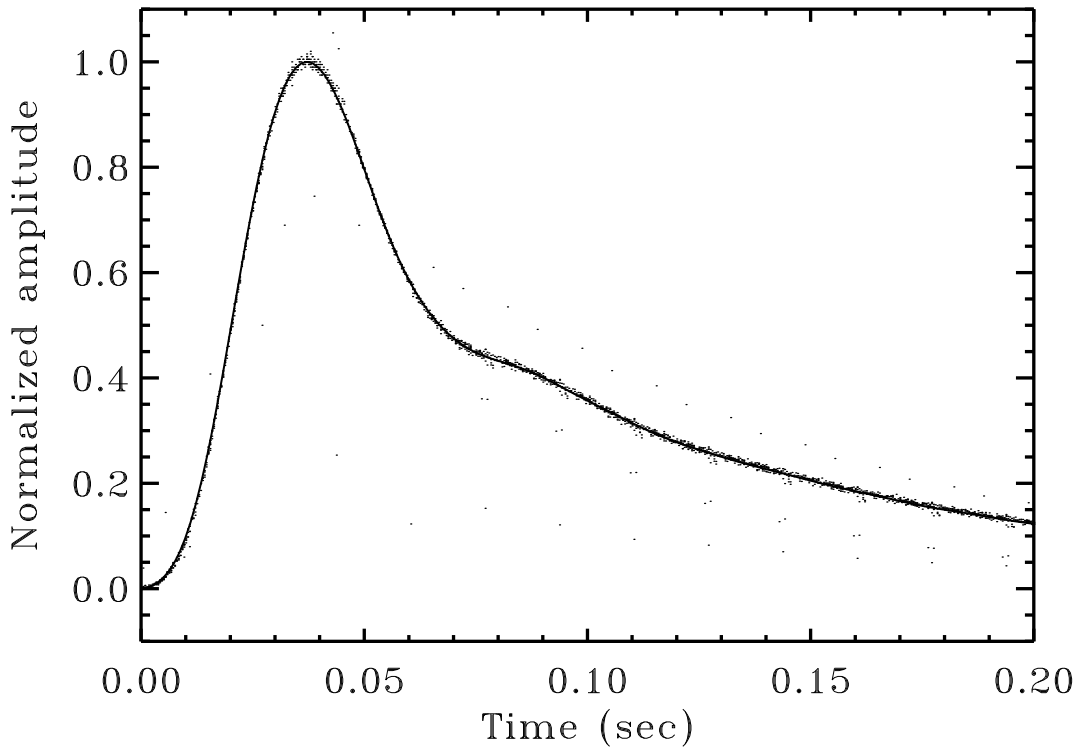}
    \caption{In-flight response of the 145W1 BOOMERanG channel to an
      impulsive event. The frequency response of the system is the
      Fourier Transform of this response. The points are accumulated
      from several cosmic-rays events shifted and normalized to fit
      the same template. (Figure extracted from \cite{masi06})}
    \label{fig:boom_tf}
\end{figure}

If an optical method is not usable for some reason, the time response
of the detector may also be estimated using the signal from cosmic ray
glitches. A cosmic-ray hit on a bolometer is well approximated by a
delta function power input. It leaves on the data-stream a typical
signature of the response of an impulsive input which correspond to
the transfer function of the detector, including electronics.  A
template of the transfer function can be obtained by piling up all
glitches in a given channel after common renormalization both in
position and amplitude. It can be either directly used as the
detector transfer function (BOOMERanG \cite{masi06}, figure~\ref{fig:boom_tf}) or used to
estimate the parameters of a model (Archeops \cite{macias07}). For
bolometers, due to the internal time constant of the detector
absorber, differences can appear with respect to the models. Indeed,
the energy deposit on the entire absorber (for millimeter-wave
photons) or in a localized area (for cosmic rays particles) affect
differently the response of the detector. In the worst case, several
other time constants can appear on glitches, depending on where the
cosmic ray hits the detector.

%%%%%%%%%%%%%%%%%%%%%%%
% FIG
%%%%%%%%%%%%%%%%%%%%%%% 

The result of a time constant is basically to low pass filter a
signal. Deconvolving the data stream by the transfer function results
in an increase of the noise at higher frequency.

\subsection{Beam}

The beams represent the optical transfer function of the instrument.
The response to point sources for many CMB experiments can often be
modeled as a 2D-gaussian, but asymmetry of beams has become one of the
most important sources of systematic problems for CMB experiments.

Beams are generally estimated using the response to a point source
such as planets or bright stellar objects, which can be combined with
physical optics models.  For symmetrical beams, a profile in one
dimension can be used to ajust the model (BOOMERanG). Otherwise, local
maps of brighter sources, such as Jupiter for WMAP \cite{page03} or
Archeops \cite{macias07}, are constructed to estimate the beam shape.

Multimode horns can show more complex beam pattern with several maxima
(for an example, see the Archeops beam pattern \cite{macias07}).

%%%%%%%%%%%%%%%%%%%%%%%
%FIG
\begin{figure}[htbp]
   \centering
    \includegraphics[height=6cm,width=8cm]{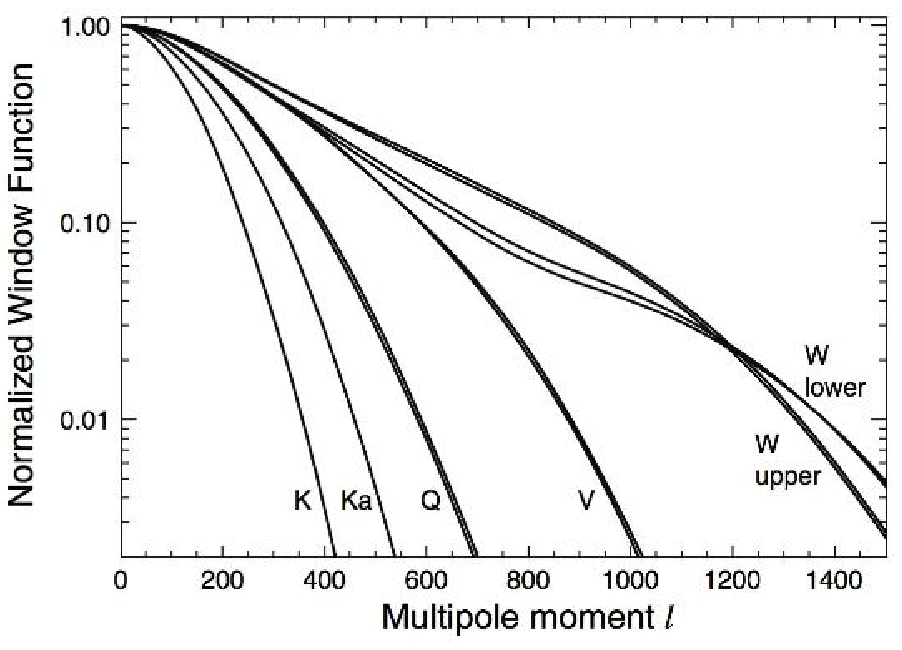}
    \includegraphics[height=6cm,width=7cm]{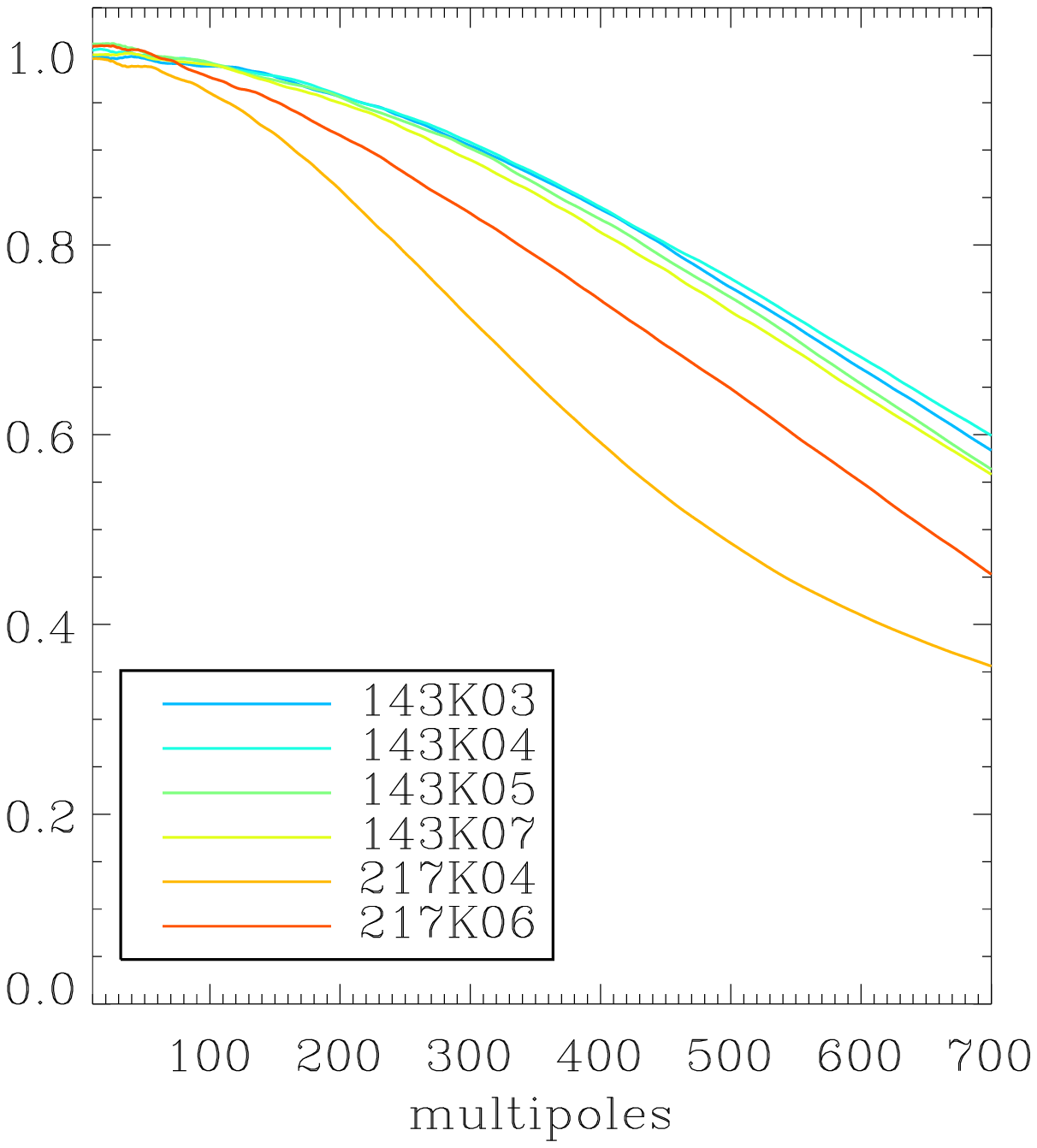}
    \caption{ Beam window function in multipole space for a WMAP ({\it
        left}, extracted from LAMBDA, http://lambda.gsfc.nasa.gov) detector, 
        and Archeops ({\it right}, extracted from \cite{tristram05b})}
    \label{fig:bell}
\end{figure}
%%%%%%%%%%%%%%%%%%%%%%% 

In CMB analyses, beam effects are often simulated and corrected on the
power spectrum rather than deconvoluted in maps domain. Simulations
include the convolution by the beam pattern so that the effect is corrected 
via a transfer function in multipoles (figure~\ref{fig:bell}).

\begin{figure}[htbp]
   \centering
    \includegraphics[angle=90,width=7cm,height=5cm]{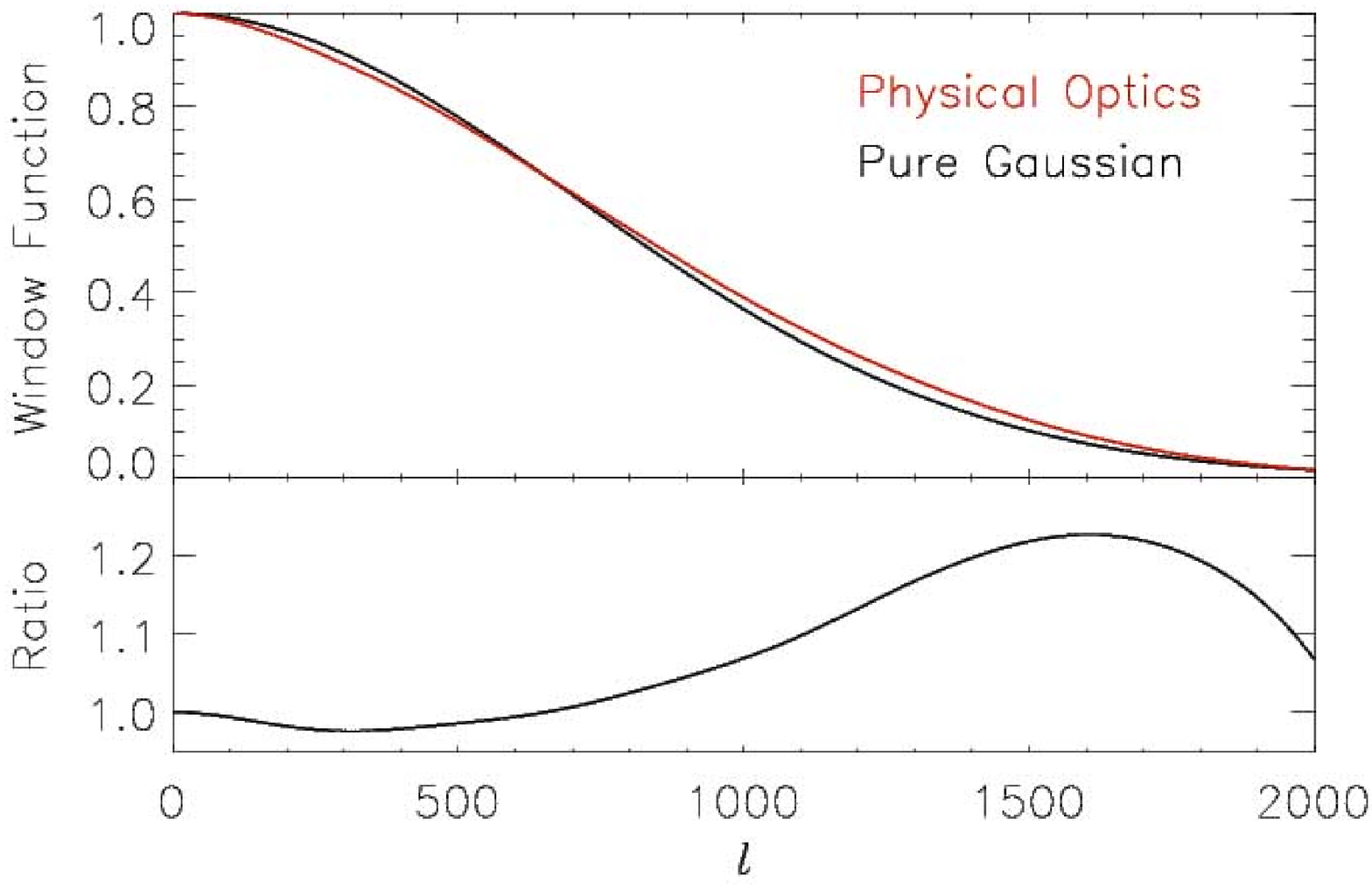}
    \hspace{0.5cm}
    \includegraphics[width=7cm,height=5cm]{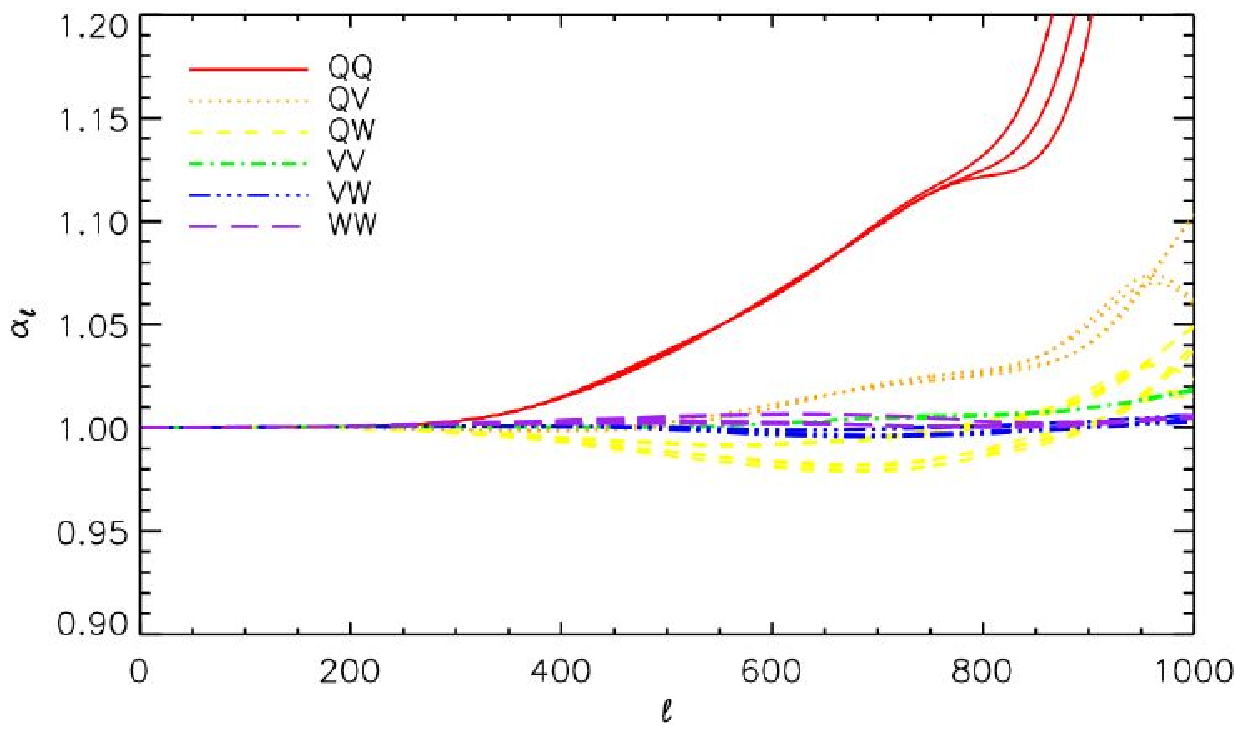}
    \caption{Effects of beam asymmetry on $C_\ell$.  {\it Left:} for a
      BOOMERanG bolometer at 145~GHz, computed from the physical
      optics model (red line) and a 9.8 arcmin FWHM gaussian beam
      (black line). (figure extracted from \cite{masi06}) {\it Right:}
      ratio between the window function for the actual beam and that
      for a gaussian beam for each WMAP channel. (Figure extracted
      from \cite{hinshaw06}) }
    \label{fig:asymetric_bell}
\end{figure}

Errors due to an asymmetrical beam pattern being treated as
symmetrical a the major source of systematics at high multipoles 
(figure~\ref{fig:asymetric_bell}).

Several different methods of modeling beam pattern have been
developed. Each observation of the sky is convolved with the beam,
which depends both on its shape and on its orientation on the sky. For
asymmetric beam patterns, the convolution is then intrinsically linked
to the pointing at each point in time, which makes it computationally
intensive. Most convolution methods work in harmonic space using
either a general convolution algorithm \cite{wandelt03} or a model of
the beam pattern in real space that can easily be decomposed in
harmonic space. For the latter, several methods have been developed in
order to symmetrize the beam \cite{page03, wu01}, to approximate the
ellipticity \cite{souradeep01, fosalba02}, or decompose the beam
pattern into a sum of gaussians \cite{tristram04} (BOOMERanG,
Archeops).

%%%%%%%%%%%%%%%%%%%%%%% 
% FIG
%%%%%%%%%%%%%%%%%%%%%%% 

\subsection{Polarization beams}

For polarization-sensitive detectors, we define the co- and
cross-polarization beams. For a given polarization sensitivity
direction at the receiver, the direction of co-polarization at the
beam center (on-axis) is conveniently defined as the image of the
sensitivity direction through the optics. The cross-polarisation
sensitivity direction is orthogonal to the co-polarization.

\begin{figure}[htbp]
   \centering
    \includegraphics[height=5.5cm]{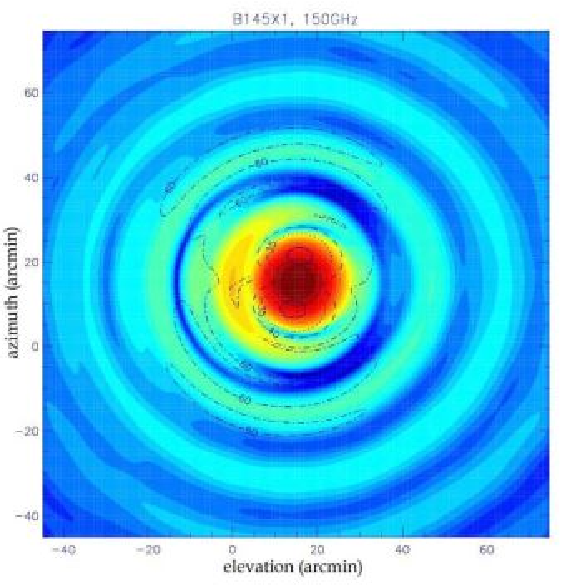}
    % \hspace{0.5cm}
    \includegraphics[height=5.5cm]{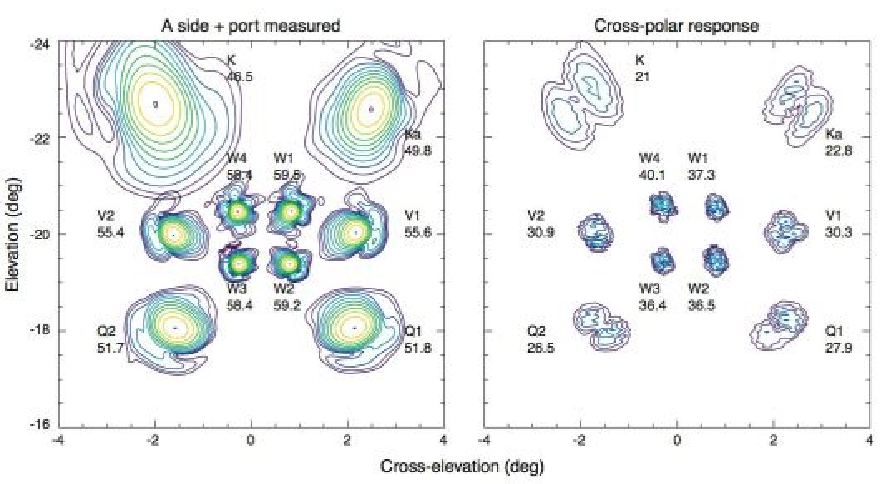}
    \caption{ {\it left:} Comparison of the cross-polar (contours) and
      co-polar (colors) beams for one of the BOOMERanG 145~GHz
      channels, as computed with the physical optics code BMAX.
      (figure extracted from \cite{masi06}) {\it right:} WMAP-measured
      focal plane for the A side for the co- and cross-polar beams.
      The contours are spaced by 3~dB and the maximum value of the
      gain in dBi is given next to selected beams. Measurements at
      twelve frequencies across each passband are combined using the
      measured radiometer response. This beam orientation is for an
      observer sitting on WMAP observing the beams as projected on the
      sky. (figure extracted from \cite{jarosik06})}
    \label{fig:xpolbeam}
\end{figure}

For data analysis, one needs to estimate the level of
cross-polarization in order to characterize the beam patterns and
reconstruct the polarized signal of the sky. Moreover, in principle,
a significant asymmetry of the main beam can contaminate the
polarization measurements.This effect depends largely on the scanning
strategy. As for the main intensity beam, the effect of the
cross-polarization on the maps is estimated using simulations.  For
experiments such as WMAP \cite{jarosik06} or BOOMERanG \cite{masi06},
any cross-polarization contribution due to the optics is negligible
with respect to the intrinsic cross-polarization of the detectors (figure~\ref{fig:xpolbeam}).

%%%%%%%%%%%%%%%%%%%%%%% 
% FIG
%%%%%%%%%%%%%%%%%%%%%%% 

\subsection{Far sidelobes}

In all radio telescopes, each beam has sidelobes, or regions of
nonzero gain away from the peak line-of-sight direction.  Due to
diffraction effects, light from regions of the sky far from the main
beam can reach the detectors.

Sidelobe response over $4\pi$~sr of sky can be measured from
ground-based sources and/or in-flight measurements of very bright
sources such as Moon or Sun \cite{barnes03}.

Sidelobe pickup introduces a spurious additive signal into the
time-ordered data for each detector. The optical systems of CMB
experiments are designed to produce minimal pickup from signals
entering the far sidelobes.  Thus systematic  artifacts remaining in
CMB maps can be based on a well-justified assumption that sidelobe
effects are small relative to the sky signal.

For most applications in radio astronomy, such weak responses would be
negligible. However, the relative brightness of Galactic foregrounds
makes side-lobe pickup a potentially significant systematic effect for
CMB measurements (3.7\% to 0.5\% of the total sky sensitivity for
WMAP \cite{page03}).

%%%%%%%%%%%%%%%%%%%%%%%%%%%%%%%%%%%%%%%%%%%%%%%%%%%%%%%%%%%%%%%%%
%  Calibration
%%%%%%%%%%%%%%%%%%%%%%%%%%%%%%%%%%%%%%%%%%%%%%%%%%%%%%%%%%%%%%%%%
\section{Calibration}

\subsection{Spectral calibration}

The power absorbed a the detector is a function of the incident
optical power, the spectral response and the optical efficiency of the
system. The spectral response of the detector is necessary for the
analysis of the data. Unlike the calibration gain and offset that can
be estimated in-flight, bandpass measurements usually must be made in
the lab.

The width of the bands are usually designed to be as broad as
possible: $\nu/\Delta\nu \simeq 3$. This gives larger bands at higher
frequencies \cite{jarosik03,benoit02,masi06}.

\subsection{Gain corrections}
The responsivity of CMB detectors depends on the loading they see.
This can evolve during the observations. To correct for the gain
variation one can make repeated measurements of a known source on the
sky (such as the CMB dipole), one can embed a calibration source
within the experiment \cite{crill03}, or one can use a model based on
housekeeping data. Gain models are non-linear functions that strongly
depend on instrument parameters (for example, detector voltages, gains
of amplifiers, phase between radiometers) and monitored temperatures.

After this correction, the calibration factor in $\rm mK/\mu\rm V$ can
be considered as constant over the flight, thus allowing for a much
easier determination.

\subsection{Absolute calibration}

Detectors measure voltage variations that are directly proportional to
the temperature variation of the sky. To get back to the temperature,
one has to determine a calibration factor by detector. The latter
could be considered constant since the time dependance has been
subtracted to first order by the linearity corrections above.

Some calibrators that can be used are: the dipole (kinetic and
orbital), the galactic diffuse emission and point sources.  Usually,
for channels dominated by the CMB (between 20 and 300 GHz),
calibration on dipole is preferred. Otherwise, at higher and lower
frequencies, galactic emission calibration can be successfully
applied. Error-bars on point source brightness temperature models and
beam model uncertainties makes this kind of calibration usually less
precise than those on diffuse emission.

\begin{itemize}
\item {\bf Dipoles}\\
  The Dipole is usually prefered for calibrating experiments with
  large sky coverage, such as COBE, FIRS, the 19~GHz Experiment,
  Archeops, WMAP and Planck. This is due to the fact that it depends
  only marginally on pointing errors, it is a stronger signal than CMB
  anisotropies by a factor 100, and it has the same electromagnetic
  spectrum, while not being so bright as to cause non-linearities in
  the detectors.

  One usually estimates the calibration factor using a linear fit of
  the time ordered data to a template containing the dipole and
  galactic emissions.  The template is made with measurements made by
  experiments such as COBE-DMR and WMAP for the kinetic dipole and SFD
  maps \cite{schlegel98} for diffuse galactic emissions.
\item {\bf Galaxy}\\
  In terms of EM-Spectrum coverage and absolute calibration, data from
  the Far Infrared Absolute Spectrophotometer (FIRAS) instrument
  on COBE \cite{mather90} are the most sensitive. 
  FIRAS products are brightness maps which are converted to photometric
  maps with the flux convention of constant $\nu I_\nu$.
  To be compared with this galactic template, maps from experiments
  need to be degraded to the FIRAS resolution of $\sim$7~degree.

\item {\bf Point sources}\\
  Point source fluxes (such as from planets) can be compared to
  brightness models. This calibration method is of particular
  importance for small coverage experiments that cannot detect the
  dipole and/or galactic emission with enough signal-to-noise.
\end{itemize}

\subsection{Intercalibration}
  CMB experiments could have large errors on absolute calibration (due
  to a small sky coverage for example). But for coadding data from
  multiple detectors, as well as for polarization measurements,
  precise intercalibration between detectors is essential. Indeed, the
  polarization signal from bolometer and radiometer experiments is
  reconstructed using differences between pairs of detectors.
  Therefore the accuracy on this reconstruction is very sensitive to
  the relative calibration.  To ensure the precision on
  intercalibration, one can compare Galactic profiles at constant
  Galactic longitudes. Relative-calibration factors (usually done on a
  per-frequency basis) are then obtained by a $\chi^2$ minimization
  that can be constrained or not via Lagrange multipliers.  For
  polarized detectors, the presence of strongly polarized regions of
  the sky, especially in the Galactic plane, may affect the
  determination of the intercalibration coefficients. To avoid this
  effect, we proceed iteratively and mask, at each step, the strong
  polarized areas using the projected maps constructed with the
  intercalibration factors.  Attention is paid to build a common mask
  for all detectors that have to be compared.

%%%%%%%%%%%%%%%%%%%%%%%%%%%%%%%%%%%%%%%%%%%%%%%%%%%%%%%%%%%%%%%%%
%  Noise properties
%%%%%%%%%%%%%%%%%%%%%%%%%%%%%%%%%%%%%%%%%%%%%%%%%%%%%%%%%%%%%%%%%
\section{Data quality checks and noise properties}

For further processing of the data, one assumes that the noise is
gaussian and piecewise stationary. Statistical tools are used to
describe and validate the treatment described above before projecting
the data into sky maps.  This can be used to check that individual
detectors have no strange behavior or inhomogeneous properties.

\subsection{Time-frequency analysis}

The power distribution of the time-ordered data in the time-frequency
(obtained using, for example, wavelets tools) can be used to find
special features in the noise in time limited domains.  These features
can be due to differences in the foregrounds signal for particular
scanning strategies at low frequencies together with $1/f$ noise of
detectors. After systematic subtraction, the power distribution should
be flat.

Time-frequency analysis, such as in \cite{maciasbourr06}, allows us to
exclude from the further processing the detectors which present either
strong or highly time variable systematic residuals.

\subsection{Noise power spectrum estimation}

Estimation of the Fourier power spectrum of the noise is essential in
CMB analysis. First, it can be used to fill the small gaps in the data
such as those due glitches or point sources subtraction, for specific
analyses and second, we need an accurate estimate of it for Monte-Carlo
purposes.

Gap filling is necessary for map making process and Fourier power
spectrum estimations that requires continuous data (for example, if we
want to use the fast fourier transform). Gaps are filled with what we
call locally constrained realizations of noise. Simple algorithms are
based on a reconstruction of low and high frequency components
separately. First, we reconstruct the low frequency noise contribution
via an interpolation within the gap using an irregularly sampled
Fourier series. Finally, we compute the noise power spectrum locally
(in time intervals around the gap) at high frequency and we produce a
random realization of this spectrum.  Notice that we are only
interested in keeping the global spectral properties of the data.
Moreover, the gaps are in general very small in time compared to the
piece of the data used for estimating the power spectrum, and
therefore this simple approach is usually accurate enough.

Both Maximum Likelihood map making and angular power spectrum estimation
can heavily depend on the knowledge of the noise spectral density.
Bayesian approaches can be used in order to estimate the noise \cite{natoli02} 
or simultaneously the noise and the signal \cite{ferreira00} in the data.
Considering the low signal-to-noise ratio in CMB data, a first estimate 
of the noise power spectrum can be directly derived from the data themselves. 
Then, we can iterate to higher precision. We found algorithms that rely 
on the iterative reconstruction of the noise by subtracting from the TOD an 
estimate of the sky signal \cite{amblard04} useful. This latter is obtained from a
coadded map which at each iteration is improved by taking into account
the noise contribution.

\subsection{Gaussianity of the noise}

To this point, we have only considered the power spectrum evolution to
define the level of stationarity of the data. To be complete in our
analysis we first have to characterize the Gaussianity of the noise
distribution and second check its time stability. 

Kolmogorov-Smirnov tests can be used to check the time evolution of
the noise of each detector.  The Kolmogorov-Smirnov significance
coefficient gives the confidence level at which the hypothesis that
the noise has been randomly drawn from a Gaussian distribution can be
accepted.  As intrinsic detector noise can usually be considered
Gaussian to a very good approximation, any changes in the distribution
function of the noise will indicate the presence of significant
deviations from systematics such as Galactic and/or atmospheric
signals, which are neither Gaussian nor stationary.

\cite{macias07} use a Kolmogorov-Smirnov test in the Fourier domain.
Working in the Fourier domain both speeds up the calculations and
isolates the noise, which dominates at intermediate and high
frequencies, from other contributions like the Galactic and/or
atmospheric signals at low frequency. Then, Kolmogorov-Smirnov
statistics under the hypothesis of a uniform distribution is applied
in consecutive time intervals, which are compared two by two.

%%%%%%%%%%%%%%%%%%%%%%%%%%%%%%%%%%%%%%%%%%%%%%%%%%%%%%%%%%%%%%%%%
%  Map Making
%%%%%%%%%%%%%%%%%%%%%%%%%%%%%%%%%%%%%%%%%%%%%%%%%%%%%%%%%%%%%%%%%
\section{Map making}

Once the data has been ``cleaned'', the time-ordered samples must be
be projected onto a pixelized map of the sky using the associated
pointing information. To each measurement in time is associated a
pixel in its pointing direction.

The most common pixelization scheme used in CMB data analysis today is
the Hierarchical Equal Area isoLatitude Pixelization, or
HEALPix\footnote{http://healpix.jpl.nasa.gov} \cite{gorski05}, in
which each pixel is exactly equal-area, and in which pixels lay on
sets of rings at constant latitude. This allows one to take advantage
of fast Fourier transforms in the analysis, when decomposing the map data into
spherical harmonics \cite{muciaccia97}.

If the experiment has sensitivity to polarization, given the
orientations of the detectors on the sky as a function of time, maps
of the $Q$ and $U$ stokes parameters are also reconstructed from the
signal.

\subsection{The Map-making problem}
Our detectors measure the temperature of the sky in a given direction
through an instrumental beam.  This is equivalent to saying that the
underlying sky is convolved with this instrumental beam.  The
time-ordered data vector, $\mathbf{d}$, may therefore be modeled as
the sum of the signal from the pixellized, convolved sky $\mathbf{T}$
and from the noise $\mathbf{n}$:
$$
\mathbf{d} = \mathbf{A} \cdot \mathbf{T} + \mathbf{n}.
$$
The pointing matrix $A$, of size $N_t \times N_p$, relates each time
sample to the corresponding pixel in the sky.  For detectors not
sensitive to polarization, $T_p$ is the temperature of the sky in the
pixel $p$ and each element of $A$ is a scalar. For polarized-sensitive
detectors, $T_p = (I, Q, U)_p$ also contains the Stokes parameter
values in the pixel $p$, so each element of the matrix $A$ is a $3
\times 3$ matrix such that
$$
d_t = I_p + Q_p cos(2\psi_t) + U_p sin(2\psi_t) + n_t,
$$
where $\psi_t$ is the angle of the detector's polarization direction,
with respect to the polarization basis in the pixel $p$, at the time
$t$.

Defined so, $A$ is very sparse. For an ideal optimal beam, it contains only one (three for
polarization-sensitive detectors) non-null values in each row, as each
time sample is sensitive to only one pixel of the convolved sky.
For an axisymmetric beam response, the "smearing" and "pointing" operations commute, and one 
can solve for the beam convolved map. However, this is not exact for an asymmetric beam because 
the latter couples to the scanning strategy. In that case, we can use more specific method
to perform the deconvolution of the beam \cite{arnau00,burigana03,armitage04}.

At this point, it is usually assumed that the noise properties are
Gaussian and piece-wise stationary (if not, more filtering and data
cleaning are usually done). Both assumptions are crucial, as they
allow major simplifications of the map-making and power spectrum
estimation problems. Namely, Gaussianity means that all the
statistical information of the noise is contained in its covariance
matrix $\mathbf{N}$. That is,
$$
N = \left< \mathbf{n} \mathbf{n}^T \right>,
$$
where the symbols $\left<~\right>$ indicate an ensemble average.

Using the stationarity assumption, the noise can also be described by
its Fourier power spectrum, leading to major simplifications of the
covariance matrix.  It implies that each stationary block of $N$ is a
symmetric Toeplitz matrix \cite{golub96}, diagonal in the Fourier
space.

Given the above, the map-making problem becomes that of finding the
best estimate, $\hat{\mathbf{T}}$, of the sky, $\mathbf{T}$, given our
data, $\mathbf{d}$, and scanning strategy, $\mathbf{A}$ \cite{stompor02}.

\subsection{A simple solution : ``coaddition''}

If the noise in the time-ordered data is ``white'', then we can make
maps in the most intuitive manner -- simply by binning data into
pixels on the sky. This is what we call ``coaddition''.
\begin{equation}
  \hat{\mathbf{T}} 
  = 
  \left[ \mathbf{A}^T \mathbf{A} \right]^{-1} \mathbf{A}^T\mathbf{d}.
  \label{eq:coadd}
\end{equation}

The operator $\mathbf{A}^T$ sums the time-ordered data into the
correct sky pixel, while $\mathbf{A}^T \mathbf{A}$ is a diagonal
matrix, with the value of each diagonal element being the number of
time-ordered samples which have fallen into the corresponding pixel --
it gives the number of samples binned into each pixel. If the noise
in each data sample is independent, that is if the noise is white, the
constructed map is optimal in terms of signal to noise ratio.

Often, however, the noise in our measurements is correlated, resulting
in pixel-to-pixel correlations in maps. Since much of the science of
the CMB depends on measuring correlations between different points on
the sky, it is necessary to characterize and account for these
correlations, which complicates our map making procedures somewhat.

\subsection{Maximum Likelihood (ML) methods}

The most general solution to the map-making problem is obtained by
maximizing the likelihood of the data given a noise
model \cite{wright96,tegmark97}.  As the noise is Gaussian and
assuming a uniform prior on the sky temperature, the likelihood reads
$$
P(\mathbf{T}|\mathbf{d}) 
\propto 
\frac{1}{\left|(2\pi)^{N_t} N\right|^{1/2}} 
e^{-\left(\mathbf{d}-\mathbf{A}\mathbf{T}\right)^T
  \mathbf{N}^{-1}
  \left(\mathbf{d}-\mathbf{A}\cdot\mathbf{T}\right)/2}
$$

Maximizing this equation with respect to each pixel of $\mathbf{T}$
leads to the generalized least-squares (GLS) equation
\begin{equation}
  \mathbf{A}^\dag \mathbf{N}^{-1} \mathbf{A}\hat{\mathbf{T}}
  = 
  \mathbf{A}^\dag \mathbf{N}^{-1} \mathbf{d}. 
  \label{eq:mapmaking}
\end{equation}

The solution to this is, of course
$$
\hat{\mathbf{T}}
=
\left(\mathbf{A}^T \mathbf{N}^{-1}\mathbf{A}\right)^{-1}
\cdot 
\mathbf{A}^T \mathbf{N}^{-1}\mathbf{d}.
$$
Note that $\hat{\mathbf{T}}$ is also the minimum variance estimate of
the map for gaussian noise. If the noise is not Gaussian, the GLS estimator still has 
the minimum variance among all linear estimators. Also, if the noise in the time-ordered data is white,
$\mathbf{N}$ is diagonal and this solution is the same as the
``coadded'' map solution given in equation~\ref{eq:coadd}.

The covariance matrix of the resulting map is 
$$
{\cal N} 
= 
\left(\mathbf{A}^T \cdot\mathbf{N}^{-1}\cdot \mathbf{A}\right)^{-1}.
$$

A couple of points:
\begin{itemize}
\item 
  ML map-making methods usually consider noise correlations on smaller
  subsets of the data \cite{natoli01}. In this case, the matrix
  $\mathbf{N}$ becomes circulant and the product
  $\mathbf{N}^{-1}\mathbf{d}$ can be computed in the frequency domain
  in which $\mathbf{N}^{-1}$ is diagonal.
\item In practice, the inversion or even the convenient calculation of
  the $N_p \times N_p$ matrix $\mathbf{A}^T\mathbf{N}^{-1}\mathbf{A}$
  is impossible for large datasets such as Archeops, WMAP or Planck.
  Thus, equation~\ref{eq:mapmaking} is usually solved using methods
  such as the preconditioned conjugate gradient (PCG) \cite{golub96}
  or FFT methods.  Iterations are repeated until the fractional
  difference between successive iterations has reached a low enough
  value (typically of the order of $10^{-6}$).
\end{itemize}

Among the codes that have been developed to solve the generalized least
squares (GLS) mapmaking equation using iterative conjugate gradient
descent and FFT techniques, most have already been applied
successfully to CMB data sets.  MapCUMBA \cite{doré01} has been used
to construct the Archeops and BOOMERanG maps.  Mirage \cite{yvon05}
has been successfully applied to Archeops data.  ROMA
\cite{degasperis05} has also been used to analize data from the last
(2003) Antarctic flight of BOOMERanG. These codes with MADmap
\cite{borrill07} have been extensively compared in the Planck
framework \cite{ashdown06b}.

\subsection{Destriping methods}

So-called ``destripers'' attempt to simplify the general map making
problem described above when the solutions above would
require too many resources. An early, basic destriper was developed
for the 19~GHz experiment \cite{boughn92}.

It has been shown \cite{janssen96} that instrumental noise, in
particular $1/f$ noise, can be represented by an uniform offset on a scan circle signal.
In the destriping approach, the noise is divided into a
low-frequency component represented by the offsets $\mathbf{x}$ 
unfolded on time-ordered data by the matrix $\mathbf{\Gamma}$ and a
white noise part $\mathbf{n}$ which is uncorrelated
$$
\mathbf{d} 
= 
\mathbf{A} \cdot \mathbf{T} + \mathbf{\Gamma}\cdot\mathbf{x} + \mathbf{n}
$$

The maximum-likelihood coefficients, $\mathbf{x}$, can be found from
the time-ordered data, $\mathbf{d}$, by solving
$$
\left( \mathbf{\Gamma}^T \mathbf{N}^{-1} \mathbf{Z} \mathbf{\Gamma} \right) 
\cdot \mathbf{x} 
= 
\mathbf{\Gamma}^T \mathbf{N}^{-1} \mathbf{Z} \cdot \mathbf{d}
$$
where
$$
\mathbf{Z} 
= 
\mathbf{I} 
- 
\mathbf{A} \left(\mathbf{A}^T  \mathbf{N}^{-1} \mathbf{A}\right)^{-1} 
\cdot \mathbf{A}^T \mathbf{N}^{-1}.
$$

The destriping technique for the CMB map-making has been investigated in
minute detail for the Planck satellite \cite{burigana97,
  delabrouille98, maino99, maino02a, keihanen04}, which we will follow
here. It exploits the fact that Planck is spinning and thus detectors
observe large circles on the sky over and over, with each circle being
observed several times (of order 60). Averaging over these circle
makes rings with higher signal-to-noise ratio.
To extract the values of the offsets, detripers use the redundancy of
the observing strategy by considering the intersections between rings.
Intersections are defined when two points from different scan circles
fall inside the same sky pixel.

\begin{figure}[htbp]
   \centering
    \includegraphics[angle=90,width=\textwidth]{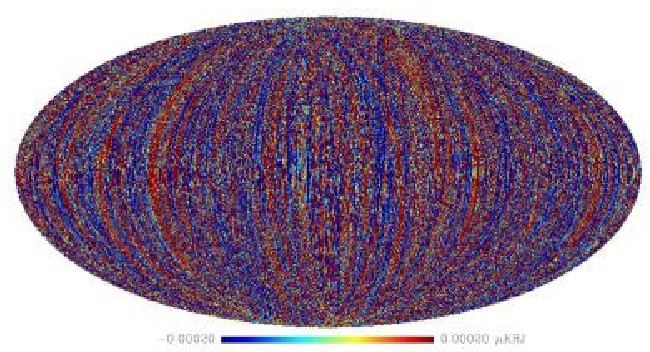}
    \hspace{0.5cm}
    \includegraphics[angle=90,width=\textwidth]{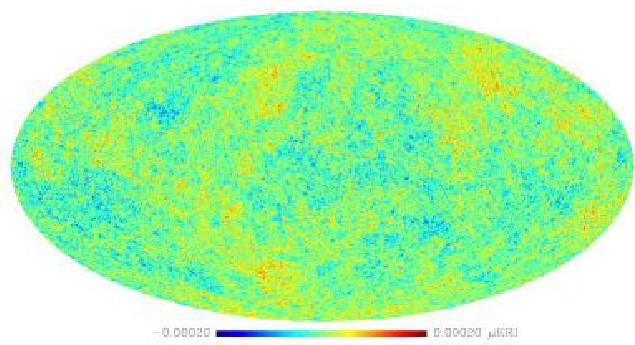}
    \caption{ Simulation at 217~GHz for Planck before ({\it top}) and
      after destriping ({\it bottom}). Simulations only include $1/f$
      noise and CMB signal.}
    \label{fig:destriping}
\end{figure}

After the amplitudes of the offsets have been estimated, they can be
used to extract an estimate of the $1/f$ noise component in the data.
A ``cleaned'' map is then obtained by simply coadding rings. Residuals
of $1/f$ noise found in the clean map for a knee frequency below
0.4~Hz have been shown significantly below the noise level
\cite{maino02a} considering a Planck-like scanning strategy 
(figure~\ref{fig:destriping}).

Destriper codes have been developed in the context of Planck
analysis~: Polar \cite{keihanen07}, MADAM \cite{keihanen05},
Springtide \cite{ashdown06a} and Polkapix \cite{perdereau07}.  In fact,
as shown in \cite{poutanen06} and \cite{ashdown06b}, the baseline length
used need not be tied to the length of a Planck ring; thus the codes
above are generalizations of the Janssen presciption. Including priors
on the low frequency
noise, the destriping algorithm is equivalent to the GLS algorithm in
the short baseline limit.  Similarly, GLS ``optimal'' codes can be
considered as destripers with a baseline fixed by the detector
sampling rate and an accurate description of the noise covariance
properties.

%%%%%%%%%%%%%%%%%%%%%%%
% FIG
%%%%%%%%%%%%%%%%%%%%%%% 

%%%%%%%%%%%%%%%%%%%%%%%%%%%%%%%%%%%%%%%%%%%%%%%%%%%%%%%%%%%%%%%%%
%  Foregrounds removal
%%%%%%%%%%%%%%%%%%%%%%%%%%%%%%%%%%%%%%%%%%%%%%%%%%%%%%%%%%%%%%%%%
\section{Foreground removal}

Maps obtained from CMB data measure the temperature variation of the
sky including astrophysical components in addition to the signal of
the CMB.  The primary goal for a CMB experiment is to remove the
foregrounds to provide a clean map of the CMB for cosmological
analysis.  But this process can also improve understanding of
foreground astrophysics.

Removal methods typically rely on the fact that the foreground signals
have quite different spatial and spectral distributions than the CMB.
Galactic free-free and synchrotron emissions decrease in amplitude
with frequency whereas Galactic dust and atmospheric emission rise
(figure~\ref{fig:foregrounds}). Thus, lower and higher frequency
channels are most sensitive to free-free and synchrotron or dust and
atmospheric emission, respectively.

\subsection{Linear combination}

Multi-frequency sky maps can be linearly combined to subtract, to
first order, Galactic signals, while preserving the CMB.  Calibration
factors and systematic effects can alter this combination when using
different data sets.

The most simple way to obtain a clean CMB map is to minimize the
variance of the resulting map, constraining the sum of the
coefficients in order to preserve the CMB signal \cite{bennett03b}.
\cite{eriksen04a} developed a similar method considerably faster by
using a Lagrange multiplier to linearize the problem. The method is
template-free and can be efficient but gives a single CMB map with
complicated noise properties.

\subsection{Template removal}

Decorrelation algorithms use templates of these emissions to subtract
parasitic signals either in the map or in the time domain. Templates
are constructed from observations, coming either from the same
experiment or not, and extrapolated in frequency.

Galactic signals dominate at low frequencies in time domain (less than a few Hz).
Templates can be bandpass filtered in that range in order to improve
the efficiency of these algorithms. The correlation coefficients are
then directly computed via a simple regression analysis. A linear
combination of the templates multiplied by the correlation
coefficients is then removed from the data.

Extrapolation in frequency and in space can give non-negligible
residuals in the decorrelated maps. It is thus important to estimate
the residual foreground uncertainties that remain after these
templates have been fit and subtracted. Nevertheless, template-based
maps have been used for the CMB anisotropy power spectra estimation of
WMAP \cite{hinshaw06}, Archeops \cite{macias07} and
BOOMERanG \cite{masi06}.

\subsection{Component separation}

Component separation algorithms take advantage of the fact that CMB
and foregrounds have different EM-spectra.

The signal from several channels at different frequencies can be
modeled as a linear combination of astrophysical components plus
noise.  This leads to a decomposed representation of:
$$
x^d_{\ell m} = A_{d c}~s^c_{\ell m} + n^d_{\ell m},
$$
where the data $x^d_{\ell m}$ are decomposed into spherical harmonic
coefficient by detector $d$. The mixing matrix $A_{d c}$ describes the
amplitudes of each of the components $c$ for each detector $d$.
$s^c_{\ell m}$ and $n^d_{\ell m}$ are the spectra of the components
and of the noise, respectively. Coefficients of the mixing matrix $A$
are linked to the electromagnetic spectrum of each component and to
the relative calibration between detectors.

Maximum likelihood methods give a simple solution of the problem in
the case where we have stationary gaussian noise characterized by its
auto-correlation matrix $N$:
$$
\tilde s = \left( A^t N^{-1} A \right)^{-1} A^t N^{-1} y.
$$

More sophisticated algorithms have been developed:
\begin{description}
\item{\bf Wiener solutions} \cite{bouchet99a, bouchet99b}. The
  Statistics of the astrophysical components must be gaussian and
  stationary and their spectrum is assumed to be known.  The
  likelihood function includes a term that take into account the
  correlation of $s$ in spherical harmonic space.
\item{\bf Maximum entropy} \cite{hobson98}.  In contrast to the Wiener
  solution, this does not assume Gaussianity of the components, but
  still requires us to know $A$ and $N$. Prior assumptions on
  components are taken into account through a additional term in the
  likelihood that contains the entropy of the system.
\item{\bf Independent Component Analysis
    (ICA)} \cite{maino02b, baccigalupi04, aumont07}.  These algorithms do not
  assume any priors on components. They compute the deviations with
  respect to a model of gaussian components equally distributed. This
  differences can result in non-gaussian effects, non-stationary
  effects and spectral dependancy effects \cite{delabrouille03}.
\end{description}

%%%%%%%%%%%%%%%%%%%%%%%%%%%%%%%%%%%%%%%%%%%%%%%%%%%%%%%%%%%%%%%%%
%  Power Spectrum Estimation
%%%%%%%%%%%%%%%%%%%%%%%%%%%%%%%%%%%%%%%%%%%%%%%%%%%%%%%%%%%%%%%%%
\section{Angular Power Spectrum Estimation}

The angular power spectrum of CMB anisotropies has become one of the
most important tools of the modern cosmology. While in the linear
regime, fluctuations predicted by most inflationary models
\cite{hu97,linde99,liddle00} result in gaussian anisotropies. In such
cases, angular power spectra both in temperature and polarization
contain all the cosmological information in the CMB. In particular,
cosmological parameters can be derived from these spectra.

In the last decade, CMB data have grown considerably in both quantity
and quality. Concurrently, methods have been developed to estimate the
angular power spectra from maps as quickly as possible. This has
allowed direct comparisons of theoretical predictions and observations
using fast and accurate statistical methods. Moreover, great efforts
have been made on simplifying angular spectra estimators so that they
can deal with the huge amount of accurate data in a reasonable amount
of time, especially in light of the new arrays of detectors that will
soon multiply the number of detectors by one or two order of
magnitude.

Except for very specific methods that, for example, estimate spectra
on rings for Planck-like scanning strategies \cite{vanleeuwen02,
  challinor02, ansari03}, most power spectra estimators belong to one
of the two following categories: maximum likelihood algorithms or
quadratic estimators usually called {\it pseudo-}estimators
(see \cite{efstathiou04} and \cite{efstathiou06} for a complete
discussion).

Maps of I, Q and U components of the CMB signal are decomposed into
spherical harmonics $\alm^T$, $\alm^E$ and $\alm^B$.  From these
coefficients, one can construct the 6 angular power spectrum~:
$C_\ell^{TT}$, $C_\ell^{EE}$, $C_\ell^{BB}$, $C_\ell^{TE}$,
$C_\ell^{TB}$ and $C_\ell^{EB}$. Systematic effects need to be taken
into account in this process. In particular, beam smoothing effects or
partial coverage of the sky must be accounted for.  Even for full sky
missions, foreground residuals usually still dominate the noise in the
Galactic plane. To avoid any contamination of the angular power
spectra, a mask is applied to suppress pixels of the sky dominated by
parasitic signal, leading to less than full-sky effective coverage.

\subsection{Maximum Likelihood methods}

Maximum likelihood algorithms \cite{bond98,tegmark97,borrill99a}
estimate angular power spectra using the angular correlation function
$M$ by maximizing the probability of $C_\ell$ considering the maps T:
$$
\mathcal{P}(C_\ell | T) 
\propto 
\exp \left[ -\frac{1}{2} \left( T^T M^{-1} T + Tr(\ln M) \right) \right].
$$

The correlation matrix of pixels $M$ includes the correlations between
pixels due to sky signal, $S$, and that due to noise $N$: $M_{pp'} =
S_{pp'} + N_{pp'}$.  Thus, the signal correlation matrix reads
$$
S_{pp'} 
= 
\sum_\ell 
\frac{2\ell + 1}{4\pi} B_\ell^2 C_\ell P_\ell\left( \cos\theta_{pp'} \right)
$$
where $\theta_{pp'}$ is the angle on the sphere between pixels $p$ and
$p'$.

Error bars are usually estimated directly from the likelihood function
which is either sampled for each multipole or approximated by a
quadratic form. The generalization to polarization can be found in
\cite{tegmark01}.

The algorithm scales as \order{N_{pix}^3} where $N_{pix}$ is the number of pixels in the map.
This implies that maximum likelihood methods are not well adapted to
surveys such as Planck which should deliver high resolution maps with
more than $10^7$ pixels \cite{borrill99b}.

\subsection{Quadratic estimator or pseudo-$C_\ell$ methods}

Contrary to maximum likelihood algorithms, {\it pseudo-}$C_\ell$
methods compute the angular power spectra directly from the data.
These spectra are biased by instrumental effects such as beam
smoothing effects, partial sky coverage or filtering of data and so
must be corrected for these effects. Methods differ in the way they
correct spectra for these effects.

An early description of this method can be found in \cite{peebles73},
and has been used for the estimation of the angular power spectrum of
clusters of Galaxies \cite{peebles74}. More recently, several methods
have been developed, among them:
\begin{itemize}
\item {\sc spice} \cite{szapudi01} and its extension to polarization \cite{chon04}: \\
  This algorithm computes the two-point correlation function $\xi$ in real space in
  order to correct for any inhomogeneous sky coverage and then integrate with a 
  Gauss-Legendre quadrature to obtain the $C_\ell$'s.

  The method uses the correlated function estimator given by
  $$D(cos\theta) = \sum_{lm} \left| a_{lm} \right|^2 \frac{1}{4\pi} P_\ell(cos\theta)$$
  The unbiased correlation function estimator can thus be obtained as
  $$\tilde\xi(cos\theta) = \frac{D_{s}(cos\theta)-\bar D_{n}(cos\theta)}{D_w(cos\theta)}$$
  where $D_{s}$ is the raw weighted pairwise estimator, $\bar D_{n}$ is the average raw noise correlation function calculated
from Monte-Carlo simulations and $D_{w}$ is the weight correlation function.

\item {\sc master} \cite{wandelt01b,hivon02}: \\
  This algorithm computes the angular power spectra directly from the
  observed maps before correcting for the inhomogeneous sky coverage
  in spherical harmonic space. An extension to polarization can be
  found in \cite{hansen03,challinor05,brown05}.
  The biased spectrum (called pseudo-spectrum) $\widetilde{C}_\ell$ rendered 
  by the direct spherical harmonics transform of a partial sky map is different from
  the full sky angular spectrum $C_\ell$ but their ensemble average are linked by~:
  $$\tilde{\VEV{C_\ell}} = \sum_{\ell'} M_{\ell\ell'} F_{\ell'} B^2_{\ell'} \VEV{C_{\ell'}} +   \tilde{\VEV{N_\ell}}.$$
  $M_{\ell\ell'}$, which is computed analytically using the spherical transform of the weight mask, 
  describes the mode-mode coupling resulting from the cut sky.
  $B_\ell$ is a window function describing the combined smoothing effects of 
  the beam and finite pixel size. $\tilde\VEV{N_{\ell}}$ is the average noise power spectrum estimated by Monte-Carlo.
  $F_\ell$ is a transfer function which models the effect of the filtering 
  applied to the data stream or to the maps.
\end{itemize}

Pseudo-$C_\ell$ estimators often make use of the fast spherical
harmonics transform that scales in \order{N_{pix}^{3/2}} for the
HEALPix pixelization scheme \cite{gorski05}. Nevertheless, they need a
precise description of the instrument (beam, filtering, noise) that
requires a large number of Monte-Carlos. These latter are also used to
estimate the power spectrum error bars.

Recently, these methods have evolved into cross-correlation methods
that deal naturally with uncorrelated noise \cite{kogut03}, and can
compute analytical estimates of the error bars \cite{tristram05a}. A
cross-correlation method derived from the {\sc MASTER} algorithm has
been used to estimate the lastest results of WMAP \cite{hinshaw03},
Archeops \cite{tristram05b} and BOOMERanG \cite{jones06,montroy06,piacentini06}.

\subsection{Hybrid methods}

Each of the previous methods make different assumptions about the data
and are sensitive to different kinds of systematic effects.
\cite{efstathiou04} and \cite{efstathiou06} show how to combine
maximum likelihood methods and pseudo-$C_\ell$ methods to take
advantage of both algorithms.
\begin{itemize}
\item{\bf high $\ell$.}  pseudo-$C_\ell$ methods, including
  cross-correlation algorithms, can estimate the $C_\ell$ quickly and
  accurately enough when instrumental noise is dominant; {\it i.e.}
  at higher multipoles. Nevertheless, approximations made in the
  correlation matrix computation imply some correlation for the lower
  points of the spectra.
\item{\bf low $\ell$.}  Maximum likelihood algorithms used on lower
  resolution maps can give very precise estimates of the spectra at
  low multipoles as well as error bars and covariance matrices.
  Nevertheless, they are very CPU consuming and are not adapted to
  high resolution maps.
\end{itemize}

\subsection{Fourier spectrum of rings $\Gamma_m$}
Several CMB experiments have performed or will perform circular scans
on the sky which we will call {\it rings} (the 19~GHz experiment,
FIRS, Archeops, WMAP, Planck).  Carrying out a one-dimensional
analysis of the CMB inhomogeneities on rings provides an alternative
to characterize its statistical properties \cite{delabrouille98}.  In
particular, some systematic effects could be easier to treat in the
time domain rather than in the two-dimensional maps; $1/f$ noise for
instance.

In \cite{ansari03}, the authors propose a scaling law that allows one
to combine spectra corresponding to different colatitude angles (e.g.
several detectors in the focal plane) before to inverting to recover
the angular power spectrum $C_\ell$.

\subsection{Gibbs samplers}

Gibbs sampling \cite{jewell04,wandelt04} allows one to sample the
power spectrum directly from the joint likelihood distribution given
the time ordered data.  While maximum likelihood methods define an
estimator to solve the {\it a posteriori} density, Gibbs sampling
methods for power spectrum estimation \cite{eriksen04b} propose
sampling parameters values $C_\ell$ from the posterior $P(C_\ell | m)$
directly (where the map vector $m$ is the least squares estimate of
the signal $s$ from the time ordered data).  The method works
iteratively by sampling the conditional distributions $P(s | C_\ell,
m)$ and $P(C_\ell | s, m) \propto P(C_\ell | s)$ \cite{tanner96}. Each
step is obtained from the previous one, drawing a random realization
from each density:
\begin{eqnarray}
  s^{i+1} & \leftarrow & P(s | C_\ell^i, m) \nonumber \\
  C_\ell^{i+1} & \leftarrow & P(C_\ell | s^{i+1}) \nonumber
\end{eqnarray}
After convergence, $P(C_\ell | m)$ is obtained by marginalization of
$P(C_\ell, s | m)$ over $s$.

The linear systems are solved using a conjugate gradient algorithm. The
choice of a good preconditioner for the linear algebra solver is of
primary importance.
Nevertheless, the calculations are dominated by the spherical harmonic
transforms that scales in ${\cal O}(N_{pix}^{3/2})$. It has been
applied to both COBE-DMR data \cite{wandelt04} and WMAP
first-year \cite{odwyer04}.

%%%%%%%%%%%%%%%%%%%%%%%%%%%%%%%%%%%%%%%%%%%%%%%%%%%%%%%%%%%%%%%%%
%  Non-Gaussianity
%%%%%%%%%%%%%%%%%%%%%%%%%%%%%%%%%%%%%%%%%%%%%%%%%%%%%%%%%%%%%%%%%

\section{Analysis of distribution of CMB anisotropies}

A consequence of the assumed flatness of the inflation potential is
that intrinsic non-linear effects during slow-roll inflation are
generally quite small, though finite and
calculable~\cite{acquaviva03,maldacena03}.  The adiabatic
perturbations arising from quantum fluctuations of the inflaton field
during inflation are thus essentially Gaussian distributed.  However,
the mechanism by which the adiabatic perturbations are generated is
not fully established. Some alternative scenarios (such as the
curvaton or inhomogeneous reheating mechanisms) can lead to a higher
levels of non-gaussianity than that found in standard, single-field
inflation. Moreover, variants of the most simple inflationary models
also predict observable levels of non-gaussianity (e.g., generalised
multi-field models~\cite{wands02}, cosmic defects~\cite{landriau03} or
late time phase transition).

Thus, to give a completely validate the inflationary fluctuation
generation mechanism, it is important to study the distribution of
phases in temperature and polarization maps and quantify the amount of
primordial non-Gaussianity present in the CMB data.  If found, a more
accurate description of this non-gausianities will be needed to
distinguish between competing models for primordial perturbation
generation.  In addition, the search for non-Gaussianities has become
a powerful tool to detect the presence of residual foregrounds,
secondary anisotropies (such as gravitational lensing,
Sunyaev-Zel'dovich effect) and unidentified systematic errors, which
leave clearly non-Gaussian imprints on the CMB-anisotropies data.

There are many techniques to test Gaussianity, many of them developed
previously as general statistical methods to test the normality of a
data set, and others specifically for CMB anisotropies.  Most
early works tested only the consistency between CMB maps and
simulated Gaussian realizations.  More recent studies now derive
constraints on a parameter, $f_{NL}$, characterizing the amplitude of
the primordial non-Gaussianity in the primordial gravitational
potential $\Phi$, characterized as a linear gaussian term plus a quadratic
contribution
$$
  \Phi(\mathbf{x}) = \Phi_G(\mathbf{x}) + f_{NL}\Phi_G^2(\mathbf{x})
$$

Among methods applied to CMB datasets we highlight:
\begin{itemize}
\item one-point moments, such as skewness and kurtosis; %(e.g. \cite{contaldi00});
\item bispectrum analyses based on the Fourier Transform of the three-point correlation function;% (applied on COBE in  \cite{ferreira98,Magueijo00});
\item geometrical estimators on the sphere;% \cite{barreiro01,monteserin05,monteserin06};
\item Minkowski functionals;% \cite{gott90,komatsu03};
\item goodness-of-fit tests;% \cite{rayner89,aliaga03,barreiro06};
\item wavelet decompositions; and% \cite{ferreira97,hobson99,barreiro00};
\item steerable filters to search for aligned structures;% \cite{wiaux05}.
\end{itemize}
Instrumental effects and observational constraints are usually taken
into account through Monte-Carlo simulations to estimate distributions
of the testing statistic and the confidence levels.

To date there is no evidence for significant cosmological
non-Gaussianity.  Several groups have claimed detections of
significant non-Gaussianities in the first-year WMAP sky
maps~\cite{tegmark03, eriksen04c, copi04, vielva04, hansen04, park04,
cruz05}.  However, the WMAP team has shown that most of these
detections are based on {\it a posteriori} statistics and can be
explained by Galactic foreground residuals, point-source residuals, or
$1/f$ noise properties~\cite{spergel06}. Further tests on WMAP 3-year
maps have found no significant deviation from Gaussianity.  

From the WMAP 3-year sky maps, the constraint on $f_{NL}$
is~\cite{spergel06}: $$-54<f_{NL}<114$$. Tests on other CMB
experiments leads to similar results: Archeops
\cite{curto06}, BOOMERanG \cite{detroia07}, MAXIMA \cite{cayon03}, VSA
\cite{savage04,rubino06}.

%%%%%%%%%%%%%%%%%%%%%%%%%%%%%%%%%%%%%%%%%%%%%%%%%%%%%%%%%%%%%%%%%
%  Cosmological parameters Estimation
%%%%%%%%%%%%%%%%%%%%%%%%%%%%%%%%%%%%%%%%%%%%%%%%%%%%%%%%%%%%%%%%%
\section{Cosmological parameters estimation}

Cosmological models, described by a given set of parameters, can
predict the shape of the CMB angular power spectra. Thus, extracting
cosmological information from CMB anisotropies means constraining the
parameters of a model given the data. Maximum likelihood is often used
as a method of parameter estimation to determine the best-fit model.
Given a class of models and an observed data set, the probability
distribution of the data (sometimes also multiplied by prior
functions) is maximized as a function of the parameters.  Then the
goodness-of-fit must be constructed in order to decide if the best-fit
model is indeed a good description of the data.  If it is, one has to
determine confidence intervals on the parameters estimation.

In principle, the likelihood function should be constructed using
pixel map values. In most of standard inflationary scenarios, CMB
fluctuations on the sky are Gaussian distributed, which implies that
pixels are random variables following a multivariate normal
distribution with a covariance matrix as a function of the model
parameters and the noise. Since the parameters enter through the
covariance matrix in a non-linear way, the likelihood function is not
a linear function of the cosmological parameters.  In practice, the
complexity of this ``full analysis'' is increased by the size of the
data set (million-pixel maps expected for Planck) and by the model
calculations \cite{bond98,bond00,borrill99c,kogut99}. Thus, the
angular power spectrum is preferred, as it reduces significantly the
size of the data set without any loss of information, for the case of
Gaussian fluctuations.

The exact meaning of the confidence intervals depends heavily on the
method used:
\begin{itemize}
\item Marginalization: based on exact prediction of the Bayes theorem.
  For a given parameter, the probability is integrate over all the
  others
  $$
  \mathcal{P}(\theta_i) = \int d\theta_{n \ne i} \mathcal{L}(\theta_n)
  $$
  It answers to the question : ``For a given value of $\theta$, how
  good a fit can we get?''
\item Minimization:
  $$
  \mathcal{P}(\theta_i) = max \left\{ \mathcal{L}(\theta_n) \right\}_{n \ne i}
  $$
  It answers to the question : ``Is there at least one good fit for this
  value of $\theta$?''
\end{itemize}

Note that these two approaches are equivalent for a gaussian
distribution of the parameters.

In both cases, finding the maximum of the likelihood surface in a
multi-dimensional space is very computationally heavy to compute.
Cosmological parameter estimators solve the problem by discretizing
the space.  For each set of parameters in a predetermined grid, they
can store either all statistical information or only the likelihood
value, $\mathcal{L}$, before marginalizing or minimizing.  Recently,
Markov Chains Monte Carlo (MCMC) likelihood analyses have become an
alternative to these gridding methods \cite{christensen01,lewis02}.
This method is based on random draws of the distribution function that
is supposed to be a ``realistic'' sample of the likelihood
hypersurface.  On this sample, one can derived the mean, variance and
confidence levels.  It allows a faster analysis: while the gridding
methods scale exponentially with the number of parameters, the MCMC
method scales linearly.

Historically, parameter estimation from CMB anisotropies started with
a small number of free parameters (less than 5), all others assumed
fixed to their nominal values. Now, the usual minimum set of parameter
is made of 5: density of baryons ($\Omega_b$), density of dark matter
($\Omega_{DM}$), density of dark energy ($\Omega_\Lambda$), amplitude
of fluctuations ($A$) and the optical depth of réionization $\tau$. To
this minimal set can be added a lot of more specific parameters such
as: the scalar index ($n_s$), the running index, dark energy equation
of state ($w$) and its derivative, the neutrino masses, etc. In
parallel, more and more observational data sets are included in the
analysis coming from Supernovae Ia, Weak lensing, Baryon Acoustic
Oscillations (BAO), Hubble constant measurement by HST, etc.  The
increase of complexity in the parameter estimation process has been
allowed by the increase of computer speed as well as by the
development of faster codes to compute the power spectra
(CMBFAST \cite{seljak96b}).

%%%%%%%%%%%%%%%%%%%%%%%%%%%%%%%%%%%%%%%%%%%%%%%%%%%%%%%%%%%%%%%%%
%  Status
%%%%%%%%%%%%%%%%%%%%%%%%%%%%%%%%%%%%%%%%%%%%%%%%%%%%%%%%%%%%%%%%%
\section{Current status of observations}

Between the first COBE/DMR CMB anisotropy detection in 1992 and the
recent WMAP detections, many sub-orbital experiments have measured the
CMB anisotropies. In temperature, with WMAP \cite{hinshaw06},
BOOMERanG \cite{jones06}, Acbar, CBI and the VSA measurements, we have
now a precise measurement of the angular power spectrum including the
third acoustic peak. For polarization measurements, spectra are just
now becoming available (from WMAP \cite{page06},
BOOMERanG \cite{piacentini06, montroy06}), DASI \cite{leitch05},
CBI \cite{readhead04} and CAPMAP \cite{barkats05}).

\subsection{Angular power spectra}

In 1992, the FIRAS instrument on the COBE satellite measured the CMB
black body spectrum at $T = 2.735\pm0.06$~K \cite{mather90} (updated
in 2002 at $T = 2.725\pm0.001$~K \cite{fixsen02}). With 7~degree
angular resolution, DMR, its sister experiment on the satellite,
constrained the low part ($\ell < 12$) of the temperature power
spectra (\cite{gorski94} and \cite{gorski96,tegmark96a}). Since then a number of experiments, both
ground-based and balloon-borne, have helped refine our knowledge of
the shape of the first acoustic peak.

%%%%%%%%%%%%%%%%%%%%%%%
% FIG
%figure des Cl
\begin{figure}[htbp]
   \centering
    \includegraphics[height=5cm,width=7cm]{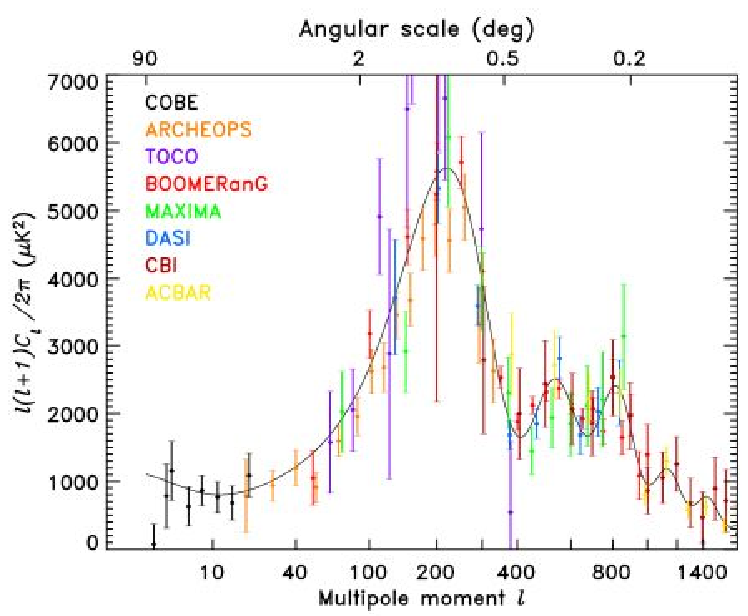}
    \hspace{0.5cm}
    \includegraphics[height=5cm,width=7cm]{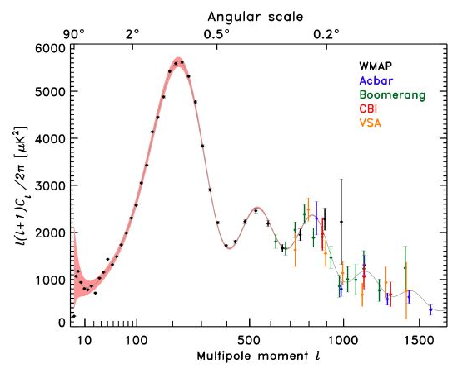}
    \caption{ 
      {\it Left:} CMB temperature angular power spectrum
      before WMAP in 2003 (COBE \cite{tegmark96a},
      Archeops \cite{benoit03}, TOCO \cite{miller02},
      BOOMERanG98 \cite{ruhl03}, Maxima01 \cite{lee01},
      DASI \cite{halverson02}, CBI \cite{pearson03} and
      ACBAR \cite{kuo04}). (extracted from \cite{hinshaw03}) \\
      {\it Right:} Results from the 3 years observation of
      WMAP \cite{hinshaw06} together with last results from high
      resolution experiments: BOOMERanG \cite{jones06},
      Acbar \cite{kuo04}, CBI \cite{readhead04} and VSA
      \cite{dickinson04}. (extracted from \cite{hinshaw06}) }
    \label{fig:cl_status}
\end{figure}
%%%%%%%%%%%%%%%%%%%%%%% 

Even before WMAP results were released, experiments when combined
clearly showed the presence of the two acoustic peak
(figure~\ref{fig:cl_status}). But the error bars were still dominated
by systematic effects, and calibration was very difficult to obtain
from a data set where the instruments, the resolution and the
frequency bands where so different. Some where dedicated to small
angular scales using a high resolution on a small sky patches, such as
DASI or CBI, whereas others, such as Archeops, covered large portions
of the sky (30\%).

In 2003, WMAP first year results \cite{bennett03a}, with improved
sensitivity and sky coverage, gave error bars dominated by cosmic
variance up to the second acoustic peak.

\begin{figure}[htbp]
   \centering
    \includegraphics[height=5cm,width=7.5cm]{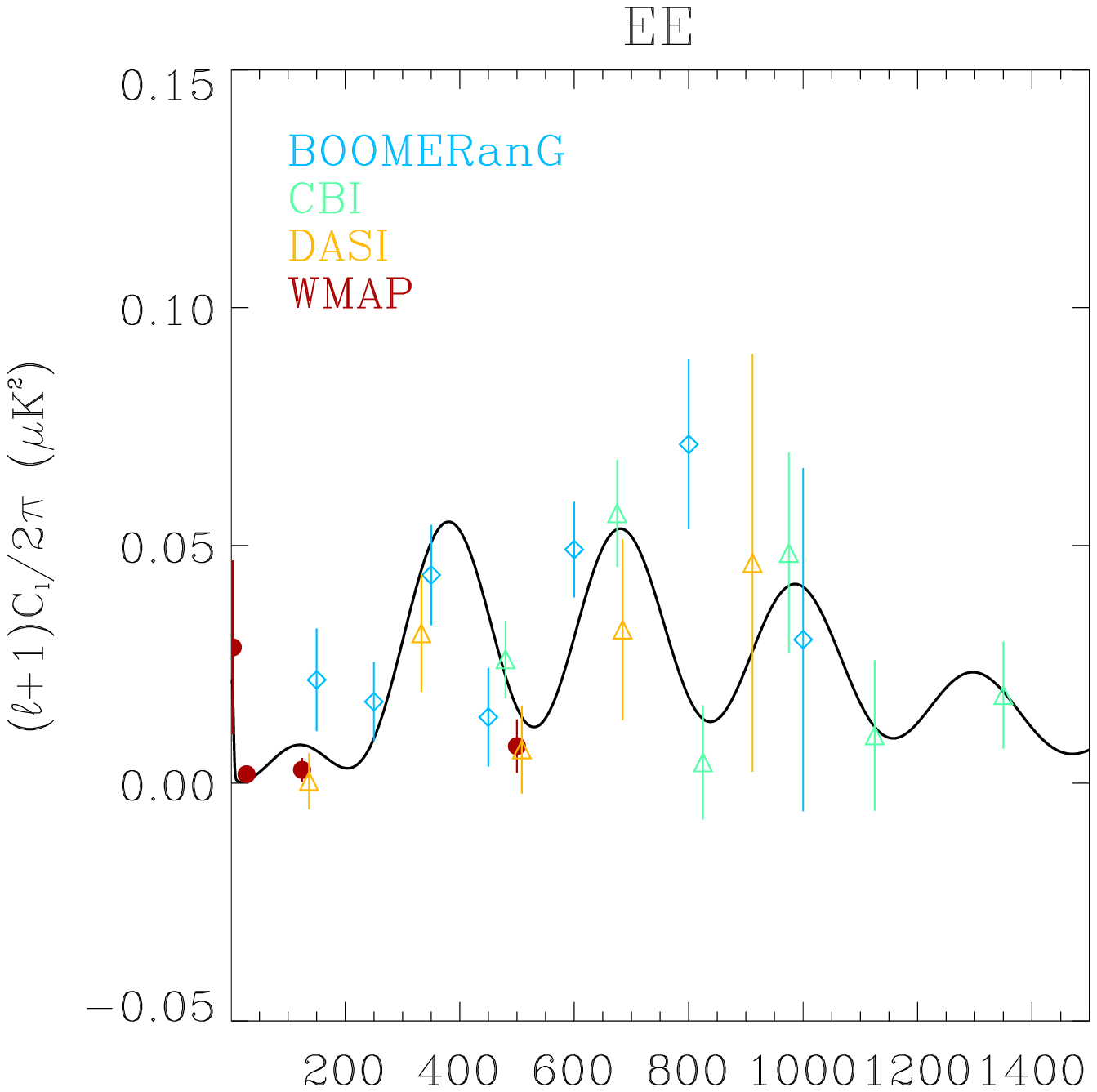}
    \includegraphics[height=5cm,width=7.5cm]{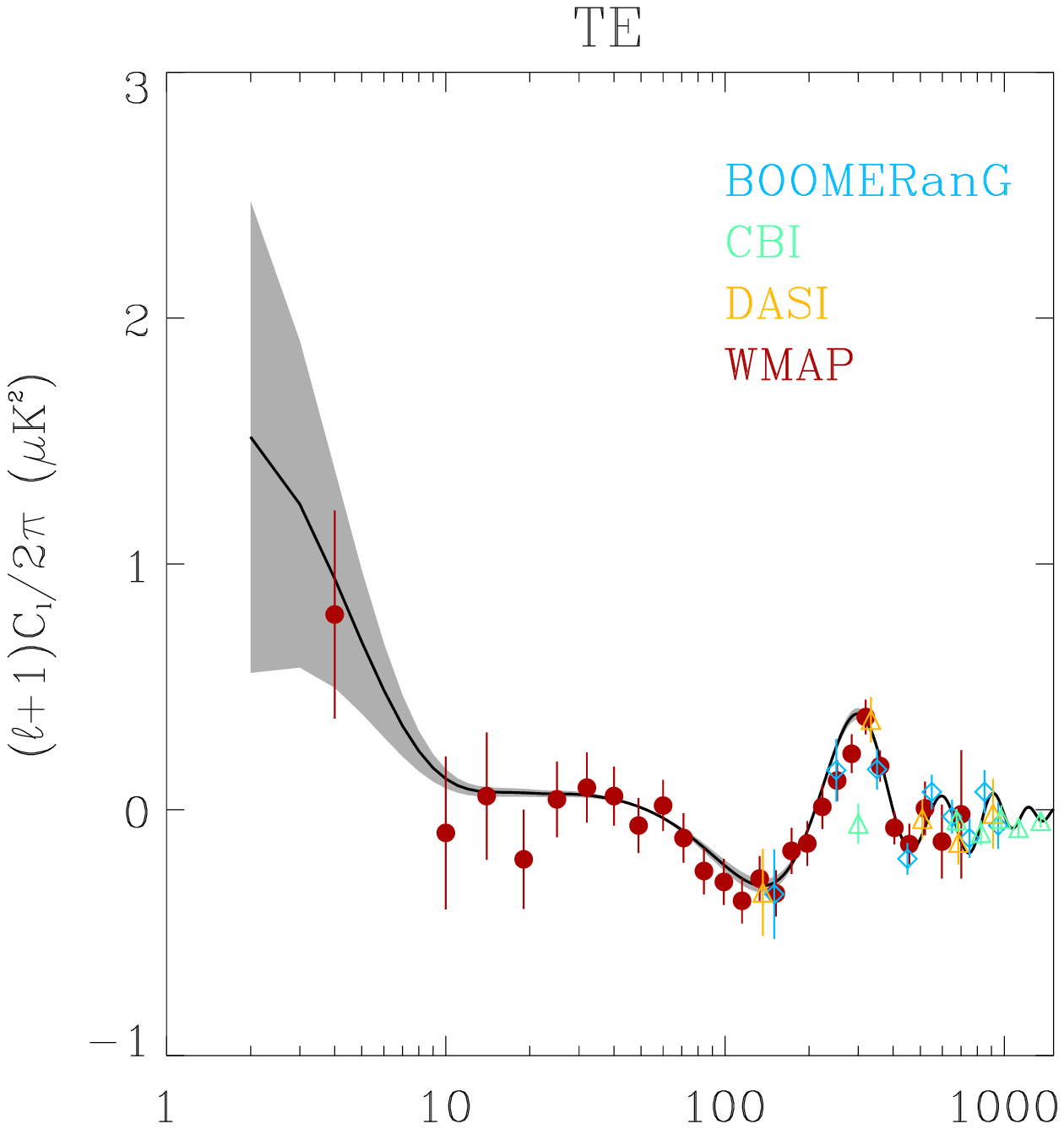}
    \caption{ E-mode spectrum ({\it left}) and TE ({\it right})
      measured by BOOMERanG \cite{montroy06,piacentini06},
      DASI \cite{leitch05}, CBI \cite{readhead04} and
      WMAP \cite{page06,hinshaw06}.}
    \label{fig:cl_polar}
\end{figure}

The first measurement of E-mode polarization was published by DASI in
2002 \cite{kovac02}, confirmed by the same team in
2004 \cite{leitch05} as well as by CBI \cite{readhead04} and
CAPMAP \cite{barkats05} (figure~\ref{fig:cl_polar}).

%%%%%%%%%%%%%%%%%%%%%%% 
% FIG
% figure des Cl TE
%%%%%%%%%%%%%%%%%%%%%%% 

We have only two measurements for the cross-correlation TE spectrum
(WMAP \cite{hinshaw06} and BOOMERanG \cite{piacentini06}). To this
point, we only have upper limits for B-mode spectrum. This illustrates
how complex CMB polarization detection is.

\subsection{Cosmological parameters}

The standard $\Lambda$CDM cosmological model agrees well with the most
recent experimental results.  Fewer than ten free parameters allow us
to fit the bulk of high precision data. Under such conditions, CMB
data put strong constraints on models with adiabatic perturbations
with close to scale invariant initial conditions, with a nearly flat
geometry and containing Dark Matter and Dark Energy. The CMB can provide
constraints on a large set of parameters and, in conjunction with other
astronomical measurements, it places significant limits on the
geometry of the universe, the nature of dark energy, and even neutrino
properties.

In \cite{spergel06}, the WMAP team fits a 6-parameter family of models
(which fixes $\Omega_{tot} = 1$ and $r = 0$), together with weak
priors (on $h$ and $\Omega_b h^2$ for example). The estimated
parameters are matter density ($\Omega_m h^2$), baryon density
($\Omega_b h^2$), Hubble Constant ($H_0$), amplitude of fluctuations
($\sigma_8$), optical depth ($\tau$), and a slope for the scalar
perturbation spectrum ($n_s$).  The fit to the 3-year WMAP data
combined with other CMB experiments yields the results in
table~\ref{tab:param_cosmo}.

%%%%%%%%%%%%%%%%%%%%%%%
%TAB
\begin{table}
\begin{center}
\begin{tabular}{c|ccc}
\hline
\hline
{Parameter} &WMAP   & WMAP& WMAP+ACBAR  \\
&Only & +CBI+VSA & +BOOMERanG \\
\hline
\hline
100$\Omega_b h^2$ & 
\ensuremath{2.233^{+ 0.072}_{- 0.091} \mbox{ }} &
\ensuremath{2.212^{+ 0.066}_{- 0.084} \mbox{ }} &
\ensuremath{2.231^{+ 0.070}_{- 0.088} \mbox{ }}  \\
$\Omega_m h^2 $ & 
\ensuremath{0.1268^{+ 0.0072}_{- 0.0095} \mbox{ }} &
\ensuremath{0.1233^{+ 0.0070}_{- 0.0086} \mbox{ }} &
\ensuremath{0.1259^{+ 0.0077}_{- 0.0095} \mbox{ }}  \\
$h$ & 
\ensuremath{0.734^{+ 0.028}_{- 0.038} \mbox{ }} &
\ensuremath{0.743^{+ 0.027}_{- 0.037} \mbox{ }} &
\ensuremath{0.739^{+ 0.028}_{- 0.038} \mbox{ }}  \\
$A$ & 
\ensuremath{0.801^{+ 0.043}_{- 0.054} \mbox{ }} &
\ensuremath{0.796^{+ 0.042}_{- 0.052} \mbox{ }} &
\ensuremath{0.798^{+ 0.046}_{- 0.054} \mbox{ }}  \\
$\tau$ & 
\ensuremath{0.088^{+ 0.028}_{- 0.034} \mbox{ }} &
\ensuremath{0.088^{+ 0.027}_{- 0.033} \mbox{ }} &
\ensuremath{0.088^{+ 0.030}_{- 0.033} \mbox{ }}  \\
$n_s$ & 
\ensuremath{0.951^{+ 0.015}_{- 0.019} \mbox{ }} &
\ensuremath{0.947^{+ 0.014}_{- 0.017} \mbox{ }} &
\ensuremath{0.951^{+ 0.015}_{- 0.020} \mbox{ }}  \\
\hline
$\sigma_8$ & 
\ensuremath{0.744^{+ 0.050}_{- 0.060} \mbox{ }} &
\ensuremath{0.722^{+ 0.043}_{- 0.053} \mbox{ }} &
\ensuremath{0.739^{+ 0.047}_{- 0.059} \mbox{ }} \\
$\Omega_m $ & 
\ensuremath{0.238^{+ 0.030}_{- 0.041} \mbox{ }} &
\ensuremath{0.226^{+ 0.026}_{- 0.036} \mbox{ }} &
\ensuremath{0.233^{+ 0.029}_{- 0.041} \mbox{ }} \\
\hline
\end{tabular}
\caption{ Cosmological parameters estimated using the 3-year WMAP data and some additional data from other CMB experiments (results extracted from \cite{spergel06}).}
\label{tab:param_cosmo}
\end{center}
\end{table}
%%%%%%%%%%%%%%%%%%%%%%%

Without spatial flatness, the CMB data alone provide only a very weak
constraint for $h$.  Inversely, including a prior on $h$ from
HST \cite{freedman01} gives the best constraint on $\Omega_{tot} =
1.003^{+0.013}_{-0.017}$, although similar results come from using
Supernova Legacy Survey data \cite{astier06} or large-scale structure
data.

The addition of other cosmological data-sets allows constraints to be
placed on further parameters.  Indeed, additional data are able to
break degeneracies that exist using CMB data alone.  For example,
considering the dark energy equation of state $w$ adds a degeneracy in
the $(w,h)$-space for low values of both parameters. This can be
broken using supernovae and large-scale structure data to yield $w=
-1.06_{-0.08}^{+0.13}$.

In addition, CMB data can put limits on parameters relevant to
particle physics models.  In particular, the CMB allows us to derive
strong constraints on the sum of neutrino masses from the neutrino
density $\Omega_\nu h^2$, assuming the usual number density of
fermions which decoupled when they were relativistic. CMB data alone
place a limit on the neutrino mass of $m < 2.0$~eV (95\%
confidence) \cite{ichikawa05}. Combining the CMB with other astrophysical 
observable (galaxy clustering, supernovae data, baryon acoustic oscillations, HST)
reduce the upper bound by roughly one order of magnitude and gives the 
strongest constraint on the sum of neutrino masses \cite{fogli06,kristiansen06,seljak06}.

%%%%%%%%%%%%%%%%%%%%%%%%%%%%%%%%%%%%%%%%%%%%%%%%%%%%%%%%%%%%%%%%%%%
%ACKNOWLEDGMENT
%%%%%%%%%%%%%%%%%%%%%%%%%%%%%%%%%%%%%%%%%%%%%%%%%%%%%%%%%%%%%%%%%%%
\ack 
We acknowledge the use of the Legacy Archive for Microwave Background
Data Analysis (LAMBDA). Support for LAMBDA is provided by the NASA
Office of Space Science. We would like to thank Professor J. Silk.

\newpage
\footnotesize
\tableofcontents

%%%%%%%%%%%%%%%%%%%%%%%%%%%%%%%%%%%%%%%%%%%%%%%%%%%%%%%%%%%%%%%%%%%
%BIBLIOGRAPHY
%%%%%%%%%%%%%%%%%%%%%%%%%%%%%%%%%%%%%%%%%%%%%%%%%%%%%%%%%%%%%%%%%%%
\include{RPPbiblio}

\end{document}

%% file: RPPbiblio.tex
%%%%%%%%%%%%%%%%%%%%%%%%%%%%%%%%%%%%%%%%%%%%%%%%%%%%%%%%%%%%%%%%%%%
% Bibliography and bibfile
%%%%%%%%%%%%%%%%%%%%%%%%%%%%%%%%%%%%%%%%%%%%%%%%%%%%%%%%%%%%%%%%%%%
\def\alc{Astroph. Lett. \& Comm.}%
          % Astrophysical Letter and Communication
\def\aip{AIP Conf. Proc.}
	%AIP Conference Proceedings
\def\aj{AJ}%
          % Astronomical Journal
\def\araa{ARA\&A}%
          % Annual Review of Astron and Astrophys
\def\aplett{Astrophys.~Lett.}%
          % Astrophysics Letters
\def\apspr{Astrophys.~Space~Phys.~Res.}%
          % Astrophysics Space Physics Research
\def\apj{ApJ}%
          % Astrophysical Journal
\def\apjl{ApJ}%
          % Astrophysical Journal, Letters
\def\apjs{ApJS}%
          % Astrophysical Journal, Supplement
\def\ao{Appl.~Opt.}%
          % Applied Optics
\def\apss{Ap\&SS}%
          % Astrophysics and Space Science
\def\aap{A\&A}%
          % Astronomy and Astrophysics
\def\aapr{A\&A~Rev.}%
          % Astronomy and Astrophysics Reviews
\def\aaps{A\&AS}%
          % Astronomy and Astrophysics, Supplement
\def\azh{AZh}%
          % Astronomicheskii Zhurnal
\def\baas{BAAS}%
          % Bulletin of the AAS
\def\bain{Bull.~Astron.~Inst.~Netherlands}%
          % Bulletin Astronomical Institute of the Netherlands
\def\fcp{Fund.~Cosmic~Phys.}%
          % Fundamental Cosmic Physics
\def\gca{Geochim.~Cosmochim.~Acta}%
          % Geochimica Cosmochimica Acta
\def\grl{Geophys.~Res.~Lett.}%
          % Geophysics Research Letters
\def\iaucirc{IAU~Circ.}%
          % IAU Cirulars
\def\jcp{J.~Chem.~Phys.}%
          % Journal of Chemical Physics
\def\jgr{J.~Geophys.~Res.}%
          % Journal of Geophysics Research
\def\jqsrt{J.~Quant.~Spec.~Radiat.~Transf.}%
          % Journal of Quantitiative Spectroscopy and Radiative Trasfer
\def\jrasc{JRASC}%
          % Journal of the RAS of Canada
\def\memsai{Mem.~Soc.~Astron.~Italiana}%
          % Mem. Societa Astronomica Italiana
\def\memras{MmRAS}%
          % Memoirs of the RAS
\def\mnras{MNRAS}%
          % Monthly Notices of the RAS
\def\nature{Nature}
	% Nature
\def\nar{New Astronomy Review}%
          % New Astronomy Review
\def\nphysa{Nucl.~Phys.~A}%
          % Nuclear Physics A
\def\physrep{Phys.~Rep.}%
          % Physics Reports
\def\planss{Planet.~Space~Sci.}%
          % Planetary Space Science
\def\plb{Phys.~Lett.~B}%
          % Physics Letters B
\def\pr{Phys.~Rept.}%
          % Physical Repport
\def\pra{Phys.~Rev.~A}%
          % Physical Review A: General Physics
\def\prb{Phys.~Rev.~B}%
          % Physical Review B: Solid State
\def\prc{Phys.~Rev.~C}%
          % Physical Review C
\def\prd{Phys.~Rev.~D}%
          % Physical Review D
\def\pre{Phys.~Rev.~E}%
          % Physical Review E
\def\prl{Phys.~Rev.~Lett.}%
          % Physical Review Letters
\def\physscr{Phys.~Scr}%
          % Physica Scripta
\def\spie{Proc.~SPIE}%
          % Proceedings of the SPIE
\def\pasp{PASP}%
          % Publications of the ASP
\def\pasj{PASJ}%
          % Publications of the ASJ
\def\qjras{QJRAS}%
          % Quarterly Journal of the RAS
\def\skytel{S\&T}%
          % Sky and Telescope
\def\solphys{Sol.~Phys.}%
          % Solar Physics
\def\sovast{Soviet~Ast.}%
          % Soviet Astronomy
\def\ssr{Space~Sci.~Rev.}%
          % Space Science Reviews
\def\zap{ZAp}%
          % Zeitschrift fuer Astrophysik
\let\astap=\aap
\let\apjlett=\apjl
\let\apjsupp=\apjs
\let\applopt=\ao

%%%%%%%%%%%%%%%%%%%%%%%%%%%%%%%%%%%%%%%%%%%%%%%%%%%%%%%%%%%%%%%%%
%  BIBLIO
%
% - Final page numbers of references are required for Reports on Progress in Physics and Physiological Measurement.
% - Up to ten authors may be given in a particular reference
% - Articles in the course of publication should include the article title and the journal of publication, if known.
% - The authors should be in the form surname (with only the first letter capitalized) followed by the initials with no periods after the initials. Authors should be separated by a comma except for the last two which should be separated by ?and? with no comma preceding it.
% - The year of publication follows the authors and is not in parentheses.
% - The journal is in italic and is abbreviated.
% - The volume number is bold; the page number is Roman. Both the initial and final page numbers should be given where possible. The final page number should be in the shortest possible form and separated from the initial page number by an en rule (--), e.g. 1203?14.
% - 
%%%%%%%%%%%%%%%%%%%%%%%%%%%%%%%%%%%%%%%%%%%%%%%%%%%%%%%%%%%%%%%%%
%\References